\documentclass[fleqn,usenatbib]{mnras}

\usepackage{newtxtext,newtxmath}

\usepackage[T1]{fontenc}

\DeclareRobustCommand{\VAN}[3]{#2}
\let\VANthebibliography\thebibliography
\def\thebibliography{\DeclareRobustCommand{\VAN}[3]{##3}\VANthebibliography}

\usepackage{graphicx}	
\usepackage{amsmath}	

\title[Spiral nature identification]{Kinematic methods of spiral nature identification: limitations and applicability}

\author[V. S. Kostiuk et al.]{
Valeria S. Kostiuk,$^{1}$\thanks{E-mail: valeriekostiuk@yandex.ru}
V. D. Zozulia,$^{2,1}$
A. A. Marchuk,$^{3,2}$
A. S. Gusev,$^{1}$
I. V. Chugunov,$^{3,1}$
\newauthor
and A. A. Smirnova$^{4}$
\\
$^{1}$Sternberg Astronomical Institute, Lomonosov Moscow State University, Universitetsky pr. 13, 119234 Moscow, Russia\\
$^{2}$Saint Petersburg State University, The Faculty of Mathematics and Mechanics, Universitetsky pr. 28, 198504 St.Petersburg, Russia\\
$^{3}$Central (Pulkovo) Astronomical Observatory, Russian Academy of Sciences, Pulkovskoye chaussee 65/1, 196140 St. Petersburg, Russia\\
$^{4}$Special Astrophysical Observatory, Russian Academy of Sciences, Nizhny Arkhyz 369167, Russia
}

\date{Accepted XXX. Received YYY; in original form ZZZ}

\pubyear{\the\year{}}

\begin{document}
\label{firstpage}
\pagerange{\pageref{firstpage}--\pageref{lastpage}}
\maketitle

\begin{abstract}
The nature of spiral arms in galaxies remains debated, with several competing formation scenarios, including density waves, kinematic spirals, dynamic spirals, and recurrent modes, that are difficult to distinguish observationally. We tested two kinematic methods --- the phase reversal method and the velocity harmonic dominance method --- using toy models and mock data from TNG50 simulations. For bar resonances, the phase reversal method is strongly orientation-dependent and yields reliable corotation estimates only when the bar is aligned with the disc minor axis. For spiral arms, the phase reversal method does not always yield reliable results and should be used in tondem with another kinematic method. The velocity harmonic dominance method is more robust: it distinguishes between density wave, kinematic, and dynamic spiral patterns, although it still shows some sensitivity to orientation. Uncertainties in the rotation curve do not strongly affect its results. We also apply both methods to the observed grand-design spiral galaxy NGC~3938 as an illustrative example; the results suggest the presence of multiple spiral modes, consistent with the recurrent modes scenario and this interpretation relies on different indirect evidence. Based on our findings, we propose a strategy for classifying spiral structure in observed galaxies that combines multiple independent tests. Applying this analysis to an extended sample will allow us to validate its predictive power and to investigate the diversity of spiral formation scenarios.
\end{abstract}




\section{Introduction}
Spiral structure of galaxies is indeed a fascinating phenomenon and perhaps the most intriguing one in galactic astronomy. Its true nature cannot leave the astrophysicist's mind at rest, compelling them to return to it again and again (see reviews by~\citealp{Dobbs_2014,Shu_2016,Sellwood_2022}). Initially, attempts were made to describe spiral structure analytically by considering the dispersion relation for movement of quasi-stationary density waves in axisymmetric disc~\citep{Toomre_1964, Lin_Shu_1964}. Further, it was analysed the transformation of the induced leading spirals into the trailing ones in differentially rotating discs by so-called swing amplification mechanism \citep{Julian_1966,Toomre_1981,Binney_2020}. Later, the focus shifted to stellar disc models, first ones were conducted in~\citet{Hohl1969,Miller1970}. In particular, spirals were considered in the framework of linear theory and global modes, both analytically~\citep{Aoki_1979, Bertin_1983, Bertin1989a, Polyachenko_2005, Jalali_2005} and in N-body simulations~\citep{Sellwood_1986, Earn_1995}. At the same time the gas response reflecting in shocks caused by the stellar spiral arms~\citep{Fujimoto_1968,Roberts_1969} was considered. Consequently, the mutual interaction of gas and stellar components in isolated simulated galaxies was examined~\citep{Wada_2011,Grand_2012,Baba_2016,Ardevol_2026}. And eventually in epoch of cosmological simulations with high spatial and temporal resolution it has been possible to address spiral issue to modelled galaxies that are as much close to real ones as possible~\citep{Palicio_2025,Ghosh_2025,Quinn_2026,Grand_2026}.   
\par
According to observational studies, the majority of massive disc galaxies in the nearby Universe have prominent spiral structure~\citep{Lintott_2011,Willett_2013} and spirals are also quite common at high redshift, at least up to $z \approx$ 2--3~\citep{Kuhn_2024, EspejoSalcedo_2025}. It is therefore reasonable to assume that the lifetime of spiral structure should last long (several galactic rotations); otherwise, we would have to conclude that we live in a specific epoch when the spiral structure of all these objects have become clearly distinguishable. Unfortunately, the pattern lifetime is a parameter that we cannot directly determine from observations; however, we can address this issue using galactic simulations. From this perspective, spiral arms are easy to maintain in both N-body and hydrodynamical simulations, but long‑lived patterns are rare \citep[one of the few papers,][]{Elmegreen_1993, Orlova_2003,Saha_2016}, because they require a very narrow range of initial conditions~\citep{Bertin1989a}.
\par
From a theoretical perspective, at least quasi‑stationary spiral arms must have a constant pattern speed $\Omega_\mathrm{p} = \mathrm{const}(R)$. In this case, the pattern represents a spiral density wave — note that the wave appears not in spatial coordinates but in the space of stellar orbit parameters. According to the refined density wave theory — the global cavity mode theory~\citep{Bertin1989a,Bertin_1989b} — stationary spirals are steady waves propagating radially between restricted regions~\citep{Lynden-Bell_1972}. These are the inner (ILR) and outer Lindblad resonances (OLR) — radii where the pattern speed $\Omega_\mathrm{p}$ equals $\Omega \mp \kappa/m$ ($\Omega$ is the disc angular velocity, $\kappa$ is the epicyclic frequency, $m$ is the number of spiral arms), and the corotation radius (CR), where $\Omega_\mathrm{p} = \Omega(R_\mathrm{CR})$.
\par
A slightly different view of the nature and lifetime of spiral arms is offered by the theory of recurrent groove modes~\citep{Sellwood_1991,Roskar_2012,Sellwood_2014,Sellwood_2019}. This scenario implies a cycle of growing and disrupting spiral modes, each having its own fixed shape, pattern speed, and number of arms. Note that in this case modes can both co‑exist and replace each other~\citep{Rautiainen_1999,Masset_1997}. Another viewpoint promotes the idea that spirals are material structures rotating at approximately the same speed as the disc ($\Omega_\mathrm{p} = \Omega (R)$) and are formed by the mechanism of swing amplification~\citep{Baba_2016, Fujii_2018, Michikoshi_2014, Michikoshi_2016,Bottema_2003}. These are the so‑called dynamic spirals, which are easily maintained in simulations but are short‑lived. According to this scenario, even long‑lived spiral arms in model galaxies eventually break apart into smaller segments and then, through differential rotation, reconnect with other fragments, creating a larger structure again (see fig.~4 and 8 of \citealp{Grand_2012a, Wada_2011}, respectively).
\par
Recent investigations of spiral structure in cosmological simulations indicate that focusing on a single spiral formation mechanism is apparently a wrong strategy~\citep{Quinn_2026,Grand_2026}. It has been clearly demonstrated that spiral arms can evolve — not only in terms of visual appearance but also in the shape of the pattern speed profiles and their median values. Therefore, over time a galaxy can evolve in such a way that at different periods it exhibits features of distinct spiral nature. Notable variations occur during bar formation or tidal interaction episodes. The latter case corresponds to the kinematic spiral scenario, where a two‑armed pattern acquires an angular velocity equal to $\Omega - \kappa/2$~\citep{Donner_1994}. The gravitational interaction aligns stellar orbits, forming an extended radial structure, which in turn rotates with a precession rate and creates a spiral pattern. The maintenance of spiral arms through this mechanism has been well studied in numerous simulations~\citep{Salo_2000,Dobbs_2010a, Oh_2015, Pettitt_2016, Pettitt_2017}, and the relevance of this scenario is doubted by no one.
\par
In contrast, the influence of the bar is more ambiguous. For example, some numerical $N$-body simulations demonstrate that the initial stages of fast bar growth are associated with prominent spirals, but later they disappear as a consequence of different dynamical mechanisms \citep{Athanassoula_2012, Fujii_2018, Chiba_2022}.
Furthermore, there are discrepancies between the results of different observational studies: some find a clear correlation between the amplitude (and torque) of the bar and the spiral arms~\citep{Block_2004}, while others, using different samples, detect no such correlation \citep{Kendall_2011}. It is also worth noting that there is no consensus on the rotation speeds of these two components, even though this characteristic determines the mechanism by which spirals form in barred galaxies. If the pattern speeds of the bar and the spirals are coupled or coincide, this is in good agreement with the density wave theory and the manifold theory of spiral arms~\citep{RomeroGomez_2006, Athanassoula_2009a, Athanassoula_2009b}.
\par
Above we discuss the problem of spiral nature mostly in the context of simulations, but although simulated objects can reflect general properties of real galaxies, they are still not perfect. They have many gaps — both those under study and those not yet explored - that could influence the interpretation of spiral arms investigation~\citep{Fujii_2011,Kwak_2025} and, most importantly, are not guaranteed to fully represent real objects. That is why the study of observed spiral galaxies also makes a significant contribution. In particular, we can look for observational evidence supporting each spiral formation scenario~\citep{Foyle_2011,Peterken_2019,Sellwood_2019MNRAS,Querejeta_2025}. Since we cannot observe spiral structure evolution directly, we need to pay attention to another feature, namely the pattern speed. There is a direct method to measure it using photometric and kinematic data~\citep{Tremaine_1984,Du_2026}; however, we cannot rely on it alone, mostly because it assumes a particular shape of the spiral angular velocity in advance~\citep{Merrifield_2006,Meidt_2008} and due to other no less important reasons~\citep{Borodina_2023}.
\par
Another approach is to consider the location of the corotation radius. For its measurement, a number of independent methods have been proposed in the literature~\citep{Buta_2009,Tamburro_2008,Hernandez_2005,Font_2011,Scarano_2013,Garcia-Burillo_1998,Marchuk_2024,Marchuk_2024c,Chugunov_2025,Pfenniger_2023,Elmegreen_1995}. A clearly located corotation radius (or radii) indicates the presence of a steady spiral mode (or modes), while a scattered distribution suggests that the spiral structure may rotate with a pattern speed decreasing with radius. In a previous study \citep{Kostiuk_2024}, we gathered CR measurements and analysed their consistency for about 100 galaxies. We demonstrated that about three quarters of the objects show discrepancies in the CR distribution. This may indicate the dominance of dynamic spirals; however, this contradiction may also be caused by other important reasons. One of them refers to the possibility of an unreliable CR location measured by one method or another.
\par
A substantial fraction of the numerous CR measurements have been obtained with methods based on kinematic data. In this paper we aim to test the reliability of two approaches for deriving the corotation resonance: phase reversal~\citep{Font_2011} and velocity harmonic dominance~\citep{Canzian_1993} methods. It is worth mentioning that the complexity of considering spiral nature in real galaxies arises from the fact that the majority of observational tests are based on density wave features alone. However, this is not enough for an exhaustive interpretation; such an analysis also requires features that can help us distinguish spiral arms belonging to distinct scenarios. For this purpose, we also aim to develop a strategy that could enable us to clarify the spiral nature using velocity field data. In Sections~\ref{sec: Canzian method} and~\ref{sec: FB method} we describe the main principles of the methods we are analysing. To test the main assumptions of these approaches, we generate mock data using the TNG50~\citep{Nelson_2019} simulations (see Section~\ref{sec: data simulation}). Besides, in this paper we apply these methods on real spiral galaxy, whose observational data are described in Section~\ref{sec: observations}. In Section~\ref{sec: results_1} we examine the reliability of the method proposed by \citet{Font_2011} for estimating the bar corotation position. In the next section, Section~\ref{sec: results_2}, we apply both methods to simple toy models and mock data from simulations to reveal features that can distinguish spirals of different formation scenarios. We further show the results of the considered methods on the example of NGC~3938 and examine spiral nature question using other independent approaches (see Section~\ref{sec: application}). In Section~\ref{sec:dicussion} we formulate the main conclusions and discuss future perspectives.

\section{Methods and Data}
\subsection{Velocity harmonic dominance method}
\label{sec: Canzian method}
In this paper, we focus on kinematic methods for identifying the nature of spiral arms. For this purpose, we adopt the method proposed by \citet{Canzian_1993} to locate the corotation resonance of a two-armed spiral. This method is based on measuring the gas velocity perturbation induced by a local linear spiral density wave \citep{Lin_Shu_1964, Shu_1973}. Following the works of \citet{Canzian_Allen_1997} and \citet{Wong_2004}, we briefly introduce the main idea for determining the nature of the spirals.
\par
The observed line-of-sight velocity map of a galaxy can be represented as a sum of harmonics:
\begin{equation}
\label{eq: v_los}
V_{\mathrm{los}}(R, \theta) = V_\mathrm{sys} + \sum_{j=1}^{n} \left[ c_j(R) \cos(j\theta) + s_j(R) \sin(j\theta) \right],
\end{equation}
where $V_\mathrm{sys}$ is the systemic velocity of the galaxy, $R$ is a galactocentric radius, $c_j(R)$ and $s_j(R)$ are the harmonic coefficients, and $\theta$ is the azimuthal angle in the disc plane measured from the major axis. In the absence of perturbations, the first cosine harmonic corresponds to the rotation curve, accounting for the inclination angle $i$ as $c_1(R) = V_c(R)\sin i$, while other terms vanish.
\par
\citet{Canzian_1993} showed that a two-armed linear spiral density wave $m=2$ \citep{Shu_1973} with velocity amplitude $V_{\mathrm{sp}}$ is reflected in a specific way in the amplitudes of the first- and third-order harmonics. For a larger number of spirals, higher-order harmonics should be considered (for details, see \citep{Canzian_1993}). \citet{Wong_2004} proposed a useful form for these terms
\begin{equation}
\label{eq: c1s1c3s3}
\begin{aligned}
\frac{c_1 - V_c}{\sin i} &= \frac{V_{\mathrm{sp}}}{2} \left( \frac{\kappa}{2\Omega} - \nu \right) \sin(2\theta_{\mathrm{sp}} - \alpha); \\
\frac{s_1}{\sin i} &= \frac{V_{\mathrm{sp}}}{2} \left( \frac{\kappa}{2\Omega} - \nu \right) \cos(2\theta_{\mathrm{sp}} - \alpha); \\
\frac{c_3}{\sin i} &= \frac{V_{\mathrm{sp}}}{2} \left( \frac{\kappa}{2\Omega} + \nu \right) \sin(2\theta_{\mathrm{sp}} + \alpha); \\
\frac{s_3}{\sin i} &= \frac{V_{\mathrm{sp}}}{2} \left( \frac{\kappa}{2\Omega} + \nu \right) \cos(2\theta_{\mathrm{sp}} + \alpha),
\end{aligned}
\end{equation}
here $\theta_{\mathrm{sp}}$ is a spiral phase~\footnote{The azimuthal angle of a given point on the spiral arm, measured from the major axis.}, $\alpha$ is a pitch angle defined as $\mathrm{cot}\alpha = R\, |\partial{\theta_{\mathrm{sp}}}/\partial{R}|$. The dimensionless frequency $\nu$ and epicyclic frequency $\kappa$ are given by:
\begin{equation}
    \nu \equiv \frac{m(\Omega_p - \Omega)}{\kappa}, \quad \kappa^2 \equiv 4\Omega^2 + R \frac{d\Omega^2}{dR}.
\end{equation}

We then introduce the parameter $p_{31}$, defined as the ratio of the amplitudes of the first- and third-order Fourier coefficients of the spiral velocity perturbation, to characterise the shape of the spiral (one- or three-armed) in the residual velocity field:
\begin{equation}
\label{eq: p_31}
    p_{31} =\sqrt{\dfrac{(c_1-V_c)^2+s_1^2}{c_3^2 + s_3^2}} = \left| \dfrac{\kappa^2 - 4\Omega(\Omega_\mathrm{p} - \Omega)}{\kappa^2 + 4\Omega(\Omega_\mathrm{p} - \Omega)} \right|.
\end{equation}

Following the methodology of \citet{Canzian_1993, Canzian_Allen_1997}, and assuming a constant pattern speed ($\Omega_{\mathrm{p}} = \mathrm{const}$), the ratio of harmonic amplitudes $p_{31}$ serves as a kinematic indicator of the corotation radius. Specifically, $p_{31} > 1$ in the region where $\Omega > \Omega_{\mathrm{p}}$ ($R < R_{\mathrm{CR}}$), and $p_{31} < 1$ where $\Omega < \Omega_{\mathrm{p}}$ ($R > R_{\mathrm{CR}}$). This implies that a distinct signature appears in the residual velocity field: the transition from a single-armed to a three-armed structure near the corotation radius.
\par
Although classical density wave theory relies on a constant pattern speed, recent observations \citep{Meidt_2008} and simulations \citep{Grand_2012, Baba_2013, Grand_2026} indicate that $\Omega_\mathrm{p}$ often exhibits radial dependence. Given that the linear theory describes the local response of the disc, we can extend the application of Eq.~\ref{eq: c1s1c3s3} by treating $\Omega_\mathrm{p}$ as a radially dependent parameter, $\Omega_\mathrm{p}(R)$. In the context of galactic dynamics, dynamic and kinematic spirals are often considered. In these cases, the $p_{31}$ and $\Omega_\mathrm{p}$ are defined as follows:

\begin{equation}
\begin{aligned}
p_{31} &= 1, \quad \mathrm{for\;dynamic\;spirals} \quad \Omega_{\mathrm{p}} = \Omega; \\
p_{31} &= \frac{\kappa+2\Omega}{|\kappa - 2\Omega|} > 1, \quad \mathrm{for\;kinematic\;spirals} \quad \Omega_{\mathrm{p}} = \Omega - \frac{\kappa}{2}.
\end{aligned}
\end{equation}

Therefore, by tracing the radial dependence of $p_{31}$, we can approximately infer the nature and origin of the observed spiral structure.
\par
To calculate the profile of the first- and third-order Fourier amplitudes (Eq.~\ref{eq: p_31}) from the line-of-sight velocity map, we proceed as follows: 1) obtain the residual velocity map defined as $V_\mathrm{res} (\theta,R) = V_\mathrm{los}(\theta,R) - V_{c}(R) \cos \theta - V_\mathrm{sys}$; 2) partition it into rings of fixed width, corresponding to twice the angular resolution; 3) apply the least squares method to fit the dependence of residual velocities on the azimuthal angle using Eq.~\ref{eq: v_los} up to $n=5$. However, a significant limitation of this approach lies in the observational degeneracy between the circular velocity $v_c$ and the first-order harmonic coefficient $c_1$. In practice, these variables are not strictly separable from line-of-sight velocity maps alone. In the following sections, we assume that $c_1$ represents the first-order cosine coefficient after subtracting the circular velocity component obtained in various ways. Nevertheless, even with this constraint, the method remains robust enough to exclude certain physical scenarios regarding the nature of the spiral structure. A detailed application of this technique to toy models and cosmological simulations (where the circular velocity is known a priori) is presented in Sec.~\ref{sec: results_2}.

\subsection{Phase reversal method}
\label{sec: FB method}
Another approach to estimating the corotation radius that we investigate is also directly connected to the kinematic data of the gaseous component. Moreover, it is based on the same assumption as the previous method: streaming motions induced by the density wave change phase in the vicinity of corotation. \citet{Font_2011} proposed to identify such behaviour of the gas near the CR by considering the sign change of the non-circular velocity component. Furthermore, the authors suggested that such a feature should occur both near the Lindblad resonances (ILR and OLR) of the bar and at its corotation resonance. 
\par
The main idea is to trace the sign reverse in the radial direction on residual velocities map $V_\mathrm{res}$, which is essentially the projection of the non‑circular component onto the line of sight. The original methodology mainly consisted of identifying pixels where the sign of the residual velocities changes. Their positions are supposed to indicate the location of the so‑called ``zero velocities'' or ``phase reversal''. By analyzing the distribution of galactocentric distances of ``zero velocity'' positions, we can identify the resonance locations, which should be associated with the peaks of the constructed histogram. We took into account the methodology described in \citet{Font_2011, Font_2014a} and implemented an algorithm for detecting the ``phase reversal'' position and estimating the corotation resonance location.
\par
The initial step involves extracting a map of $V_\mathrm{res}$ from the observed velocity field. To determine the positions of ``zero velocity'', the map is divided into sectors with equal central angles. For each sector, the radial distribution of velocities resembles a scattering of points, the number of which increases with distance. To analyze the behavior of non-circular velocities as a function of radius for each polar angle, we apply a smoothed average. The step size used in the smoothing process is equal to the angular resolution of the velocity field data. Next, for each "radial slice", we record the distances at which the velocity value changes sign. Note that we mark a point as ``phase reversal'' location only under the following circumstances. First, the absolute velocity values in the inner and outer parts must both exceed the uncertainty of the observed velocity. Second, both values should be greater than the standard deviation determined at the smoothing stage of the scattered points. This approach allows us to consider only the sign changes associated with the global distribution and exclude those connected to local processes.
\par
\begin{figure*}
\includegraphics[scale=0.12]{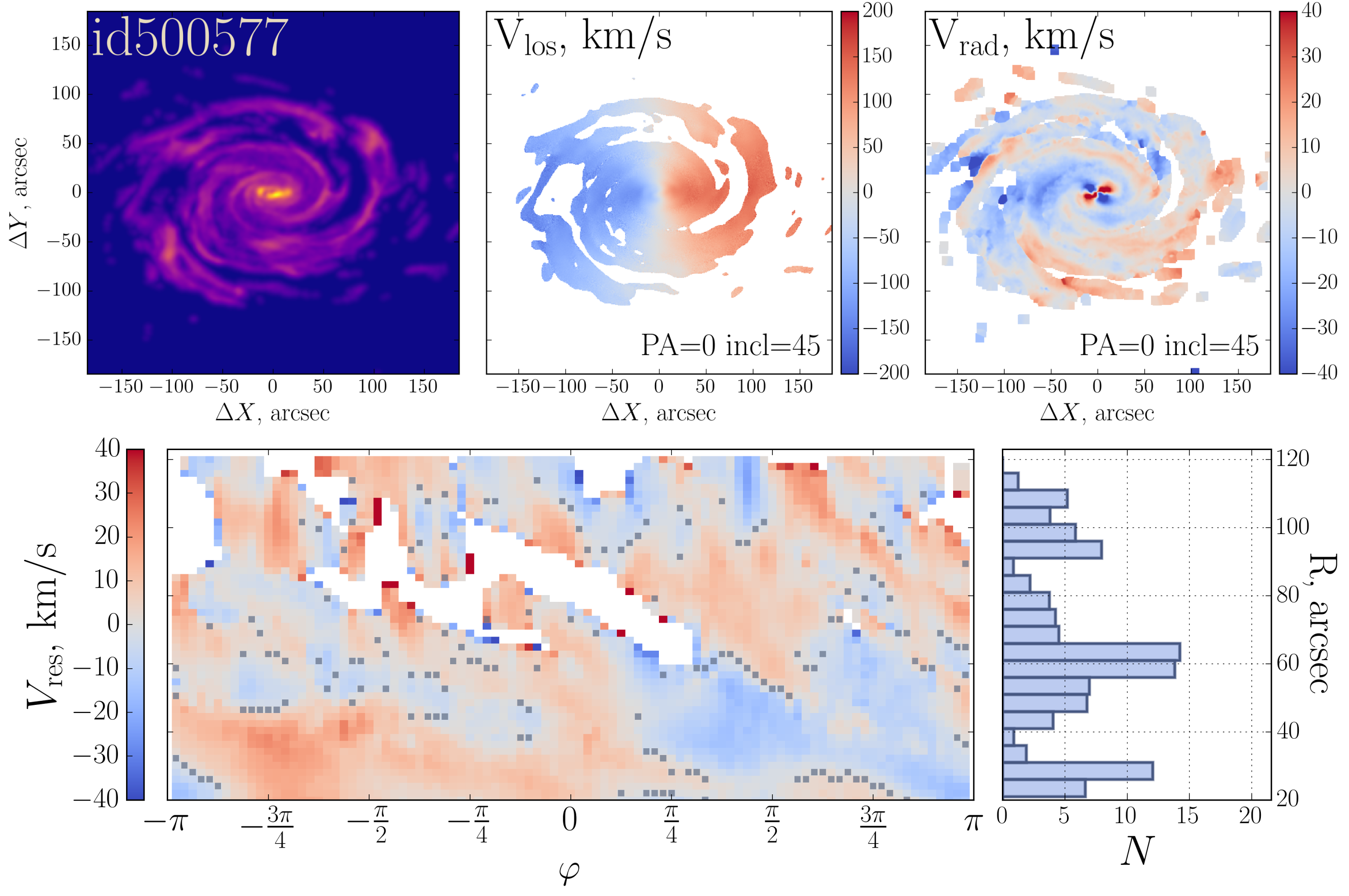}
\caption{Top row: mock data for the barred galaxy model id500577 — H$\alpha$ distribution (left), line-of-sight velocity field with $V_\mathrm{sys}=0$ (middle), and radial velocity map $V_\mathrm{rad}$ (right). All distributions are shown for an inclination of $45^\circ$ and for a bar orientation aligned with the disc major axis. Bottom panel (left): residual velocity distribution in polar coordinates, with grey squares marking the``zero velocity'' positions. Bottom panel (right): histogram of the ``zero velocity'' distances.}
 \label{fig1}
\end{figure*}

It is worth noting that we chose the smoothing step to match the angular resolution, and the final ``radial slice'' represents an array of velocities with a radial step equal to half the angular resolution. Thus, by comparing the signs of the velocity at radii separated by one step (i.e., i-1 and i+1 in the array), we ensure that the areas under comparison are separated by exactly one angular resolution. The described procedure is equivalent to detecting velocity sign changes on the residual map in polar coordinates. The bottom panel of Figure~\ref{fig1} visually demonstrates how this algorithm works; on the left of the bottom panel, there is a map of residual velocity in polar coordinates with grey squares marking the ``zero velocity'' positions. The histogram located to the right of this map shows the distribution of the distances at which the velocity changes sign.It is worth mentioning that in Appendix~\ref{sec: initial assumption} we demonstrated that the radial velocity component is better for identifying resonance locations than the non-circular one, which varies strongly with orientation. To trace the sign change of the radial velocity alone in observations, we propose to add weights to each ``zero velocity'' point equal to the sine of the polar angle during the construction of the histogram. This is why, on the bottom row of Figure~\ref{fig1}, the number of points on the residual velocity map (left) may not coincide with the value in the histogram (right) at a given radius.   
\par
\citet{Beckman_2018} demonstrated the consistency of the results obtained with this approach using both molecular and ionised gas kinematics, as well as their agreement with independent measurements from the Tremaine–Weinberg method. Moreover, the reliability of this method was confirmed by the authors in \citet{Font_2014a} by measuring the positions of bar resonances using data from a simulated barred galaxy. According to their test, the method is able to estimate the locations of the corotation radius and inner and outer Lindblad resonances. We aim to conduct the same but more extensive analysis using three barred galaxies from TNG simulation (see Section~\ref{sec: results_1}).

\subsection{Generation of the mock data}

\label{sec: data simulation}
Although the methods presented in the previous sections have been actively applied to observed galaxies, they have not been exhaustively tested on simulated data to confirm their applicability. Therefore, before interpreting their results using real data, we aim to test them on velocity fields extracted from the TNG50 simulations. This analysis allows us to examine the sensitivity of the methods to changes in the orientation of the galactic disc. 
\par
In addition, we are able to determine the precise locations of bar resonances in the simulations, which allows us to assess the validity of the method by comparing its results with the true resonance positions. A further benefit of this analysis concerns the circular velocity profile. Because deriving the true rotation curve ($V_\mathrm{rot}$) from an observed velocity field is not straightforward, the use of simulated data helps us to understand how accurately we can obtain this profile. 
\par
As in this research we aim to apply considered above methods on velocity fields that were obtained with the Fabry-Perot Interferometer (FPI) attached to the 6-m aperture telescope BTA of Special Astrophysical Observatory of Russian Academy of Science (SAO RAS)~\citep{Moiseev_2021}, we transformed the simulated data into observed ones based on the main features of this instrument (see Section~\ref{sec: observations}, for details).
\par
To generate observed H$\alpha$ velocity maps for the TNG50 galactic models~\citep{Nelson_2019}, we created H$\alpha$ image of $512\times512$ pixels with a pixel scale of $0.72''$ and applied Gaussian smoothing to account for an atmospheric seeing of $2''$. We assumed a distance to the galaxy of 20 or 30 Mpc to ensure it remains almost entirely within the "field of view". Since the cosmological simulation does not contain the information about the ionised hydrogen, we employed the canonical relation between star formation rate and H$\alpha$ luminosity from \citet{Kennicutt_1998, Pillepich_etal_2019}:
\begin{equation}
    \mathrm{SFR}\;[\mathrm{M_\odot\;year^{-1}}] = 7.9 \cdot 10^{-42}  L(\mathrm{H}\alpha) \; [\mathrm{ergs\; s^{-1}}].
\end{equation}
The final distribution of the ionised hydrogen luminosity is presented on the left top panel on Figure~\ref{fig1}. However, note that the resulting image looks quite different from real H$\alpha$ maps of galaxies (compare with Figure~\ref{fig2}): simulation does not reproduce small-scale clumpiness \citep[the reason is insufficient resolution of simulation, see e.g.][]{Ramesh_2024}. We then estimated the signal-to-noise ratio ($S/N$) and the corresponding velocity error following \citep{Moiseev_2015}\footnote{The detector noise is assumed to be low. The brightness of the sky background is set to $7\cdot10^{-18}$ erg s$^{-1}$ cm$^{-2}$ \AA$^{-1}$ arcsec$^{-2}$. The characteristic filter width is 56 \AA, and the exposure time is approximately 6000 s.}. Thus, the line-of-sight velocity in each pixel is set to zero if $S/N < 3$. Otherwise, it is defined as the weighted average velocity of the gas cells --- where the weight is proportional to $SFR$ or $L(H\alpha)$ --- plus a random variable drawn from a Gaussian distribution with a variance equal to the velocity error (see top central panel on Figure~\ref{fig1}).
\par
For additional verification of the assumptions about the gas flows in the model galaxy, which are implied in methods, we also investigate the distribution of true non-circular velocities (see Appendix~\ref{sec: initial assumption}, for details). To this end, we plot the true radial velocity map $V_\mathrm{rad}$ and the true observed line-of-sight velocity composed of radial and non-circular azimuthal components (excluding the $z$ velocity component). These maps have the same resolution as the observed ones. Note that the coverage of the radial velocity distribution (Figure~\ref{fig1}, top left) is noticeably larger than that of the line-of-sight map (Figure~\ref{fig1}, top centre). This happens because $V_\mathrm{los}$is the ``observed'' map, with coverage determined by the signal-to-noise distribution, while $V_\mathrm{rad}$ is a physical map with coverage determined by the presence of gas particles in a given area. The velocities are averaged over all gas cells within the pixel with a weight proportional to the SFR. To calculate non-circular azimuthal components, we obtain the true rotation curve, which is derived directly from the model potential calculated via the \textsc{agama} package \citep{agama}.

\subsection{Observational data}
\label{sec: observations}
\begin{figure*}
\includegraphics[scale=0.285]{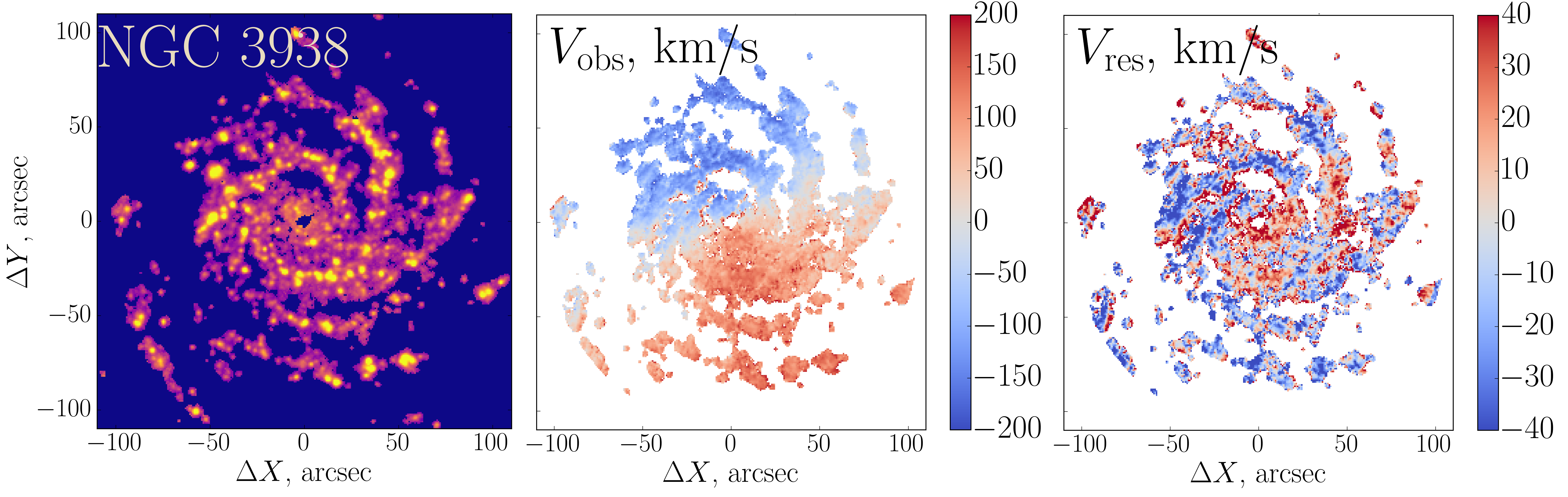}
\caption{Left and middle panels show the observed H$\alpha$ luminosity distribution and the line-of-sight velocity field corrected for the systemic velocity ($V_\mathrm{sys} = 798$~km/s). The right panel shows the residual velocity map $V_\mathrm{res}$. }
 \label{fig2}
\end{figure*}
In Section~\ref{sec: application} we present an example of spiral nature analysis on grand design spiral galaxy NGC~3938 using considered in this paper kinematic methods. This object is located at 17.9 Mpc~\citep{Palicio_2025} and having with inclination and position angles equal to 15$^\circ$ and 27$^\circ$, respectively~\citep{Abdeen_2020}.

Observations with the scanning Fabry-Perot interferometer (FPI) were performed on May 4/5, 2025, with the SCORPIO-2 multi-mode focal reducer \citep{scorpio2} installed in the prime focus of the BTA 6-m telescope of SAO RAS. During the observations, 40 interferograms with the different gap between the FPI plates were consecutively obtained. The narrow-band filter with FWHM $\sim 14$~\AA\ centered on the $\lambda=6599$~\AA\ was used to cut the operating spectral range around the H$\alpha$ line. The total exposure time was 4800~s, with a seeing of $1.3''$. The spectral sampling was 0.90~\AA. The detector, the E2V 261-84 CCD camera, provided a field of view of about $6.7'$ with a scale of $0.79''$/px in the readout mode with the $4\times4$ hardware binning.

After the primary reduction, sky subtraction, photometric and seeing corrections \citep[a detailed description of the software used and references to original papers see in][]{Moiseev_2021}, the observations were presented as a data cube, in which each pixel in the field of view contained the spectrum of the region around the H$\alpha$ line. The resulting angular resolution after smoothing during the primary reduction is $1.9''$, and the spectral resolution is $1.8$~\AA\ ($82.5$~km\,s$^{-1}$).

Following \citet{Moiseev2008}, we fitted a Voigt function to the data cube to obtain 2D distributions of the H$\alpha$ flux, the radial velocity field in this line, and the field of the radial velocity dispersion. The maps of H$\alpha$ emission and radial velocity field are shown on the left and central panels of Fig.~\ref{fig2}, respectively. The right panel on Fig.~\ref{fig2} demonstrates the distribution of residual velocity extracted using the rotational curve obtained by \textsc{Barolo} software at fixed PA and inclination with radius.

\section{Results}
\subsection{Phase reversal method: bar resonances}

\label{sec: results_1}
In this section, we test whether the phase reversal method can capture the kinematic features that occur near the main resonances, specifically the corotation one. As it was mentioned before, one of the advantages of simulated data is the possibility of accurately estimating the positions of resonances associated with the bar structure, in contrast to spiral arms. To extract the bar pattern speed $\Omega_\mathrm{bar}$ and its accuracy, we use the Python script \texttt{patternSpeed.py} from \citet{Dehnen_2023}. This script is also able to estimate the bar length in any N-body simulation. Then, using the angular velocity profile of the disc $\Omega(R) = V_\mathrm{rot}(R)/R$ (obtained from the \textsc{agama} rotation curve), we determine the radii at which the bar pattern speed $\Omega_{bar}$ is equal to $\Omega$, $\Omega+\kappa/4$ and $\Omega+\kappa/2$, corresponding to the corotation, outer ultraharmonic, and outer Lindblad resonances. This allows us to check the validity of the method for determining at least the corotation position of the bar. Note that such a validation has already been carried out in~\citet{Font_2014a}; however, our more extensive analysis enables us to identify additional restrictions and caveats.
\par
Being able to extract the true distribution of non-circular velocity from simulations, we can verify whether the detection of phase shifts on this map can reveal the positions of corotation and other resonances of the bar. In Appendix~\ref{sec: initial assumption}, we apply the ``phase reversal'' detection method to maps of non-circular velocity at various orientations, including the face-on view, for a barred galaxy. Based on this analysis, we conclude that using only non-circular velocities can lead to incorrect results due to a strong dependence on the orientation of the disc. In contrast, the same procedure applied to radial velocities alone shows a much more expected consistency between the distribution of ``zero velocity'' positions and the corotation radius, with only weak sensitivity to orientation. Therefore, we propose to focus on the radial component rather than the non-circular one.
\par
Extracting the radial velocity component from an observed velocity field is not trivial and is an almost unsolvable problem without additional data or assumptions. Since the method requires only the sign of the velocity rather than its exact value, the residual velocity map can be used for this purpose. Note that the contribution of the radial component to the non-circular velocity is maximum along the minor axis and almost disappears in the vicinity of the major one. To minimise the contribution from the non-circular azimuthal component, we assign a weight to each point where the residual velocity changes sign, equal to the sine of the polar angle at that location.
\par
\begin{figure*}
\includegraphics[scale=0.11]{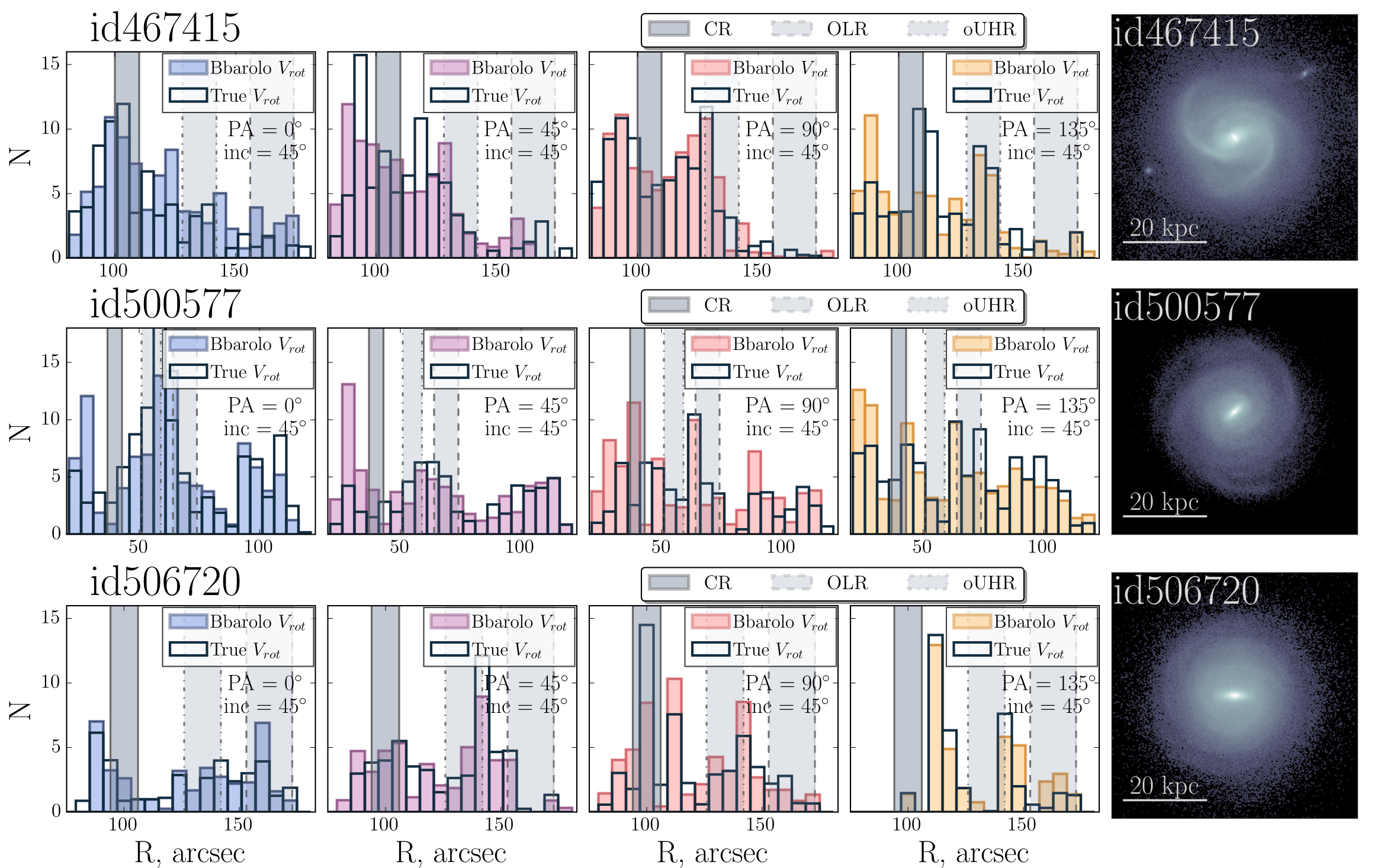}
\caption{Results of the phase reversal method for three galaxies from the TNG50 simulations: id467415 (top row), id500577 (middle row), and id506720 (bottom row). Each column shows the radial distributions of ``phase reversal'' locations for different bar orientations in the disc plane. The coloured histograms were obtained using the \textsc{Barolo} rotation curve, while the black histograms correspond to the true $V_\mathrm{rot}$ calculated via the \textsc{agama} package. Shaded vertical bands indicate the true positions of the main bar resonances (see legend in the top right of each row). To the right of each row, the stellar distributions of the corresponding galaxies at the face-on orientation are presented.} 
 \label{fig3}
\end{figure*}

The histograms presented in each row of Figure~\ref{fig3} show the distributions of ``zero velocity'' locations obtained from the velocity fields of three barred galaxies. Each column corresponds to a different orientation of the major axis at a fixed inclination angle. $\mathrm{PA} = 0^\circ$ and $\mathrm{PA} = 90^\circ$ are the directions at which the major axes of the bar and the disc are aligned with and perpendicular to each other, respectively. To avoid additional uncertainties in the resulting data, we adopt the initial values of the position angle and inclination and keep them fixed across all radii.
\par
The coloured histograms correspond to the residual velocity field obtained with the \textsc{Barolo} (2D) tool~\footnote{\hyperlink{https://bbarolo.readthedocs.io/en/latest/}{https://bbarolo.readthedocs.io/en/latest/}}~\citep{Teodoro_2015,Teodoro_2021}. It is worth emphasising that \textsc{Barolo} provides a rotation curve that can differ from the true one. To validate whether the use of non-pure circular velocities causes significant discrepancies, we also applied the same approach to residual velocity maps obtained using the true rotation curve (see histograms outlined in black). In addition, in each panel the resonance locations are marked by shaded areas outlined with solid (CR), dashed (OLR), and dash-dotted (oUHR~\footnote{Outer ultraharmonic resonance;  this position is calculated as radius where pattern speed $\Omega_\mathrm{p}$ is equal to $\Omega+\kappa/4$}) lines. The uncertainties in their positions are estimated using the errors in $\Omega_\mathrm{p}$ from \texttt{patternSpeed.py} \citep{Dehnen_2023}.

According to the distributions for all galaxies, in most cases the coloured histogram peaks do not coincide with the bar corotation radius. A similar discrepancy is present even when the true rotation curve is used (black histograms in Fig.~\ref{fig3}), indicating that the shift is not solely caused by inaccuracies in the extracted rotation curve. This discrepancy likely arises because we focus primarily on regions near the minor axis, making the resulting histogram dependent on the polar angles where the gas changes its radial direction. Additionally, the azimuthal velocity component may also affect the result, particularly near strong spirals, where this component can dominate and distort the radial velocity distribution even close to the minor axis. A further contribution may come from a physical dependence of the line-of-sight gas kinematics on the disc orientation, which would shift the apparent resonance radii. The comparison between the black and coloured histograms, in turn, points to another reason for the discrepancy. The additional shift between them is strongest at the PA values and radii where the \textsc{Barolo} rotation curve deviates most from the true one (see bottom panel of Fig.~\ref{apfig2}). Although the rotation curve uncertainty may shift the peaks less strongly than the other effects discussed above, it should still be taken into account.
\par 
From Fig.~\ref{apfig2} we can track where the deviation of the rotation curve from the true one becomes large enough to produce a noticeable shift of the peak. The rotation curve obtained with \textsc{Barolo} deviates more from the true one the stronger and more prominent the spiral arms of a given galaxy are. The ``observed'' $V_\mathrm{rot} (R)$ deviates more from the true one at radii where the spiral arm crosses the major axis, or conversely where a low-intensity region is encountered. Furthermore, Fig.~\ref{apfig2} clearly demonstrates that both the \textsc{Barolo} rotation curve and the phase-reversal method, respectively, cannot be trusted inside the bar region (hatched area in the bottom row). This happens because the gas trajectories become more elliptical within the bar region. Moreover, we cannot exclude the contribution of the vertical velocity component, which can be minimised only for galaxies with a dynamically cold gaseous disc and/or by considering highly inclined galaxies.
\par
From a theoretical perspective, the velocity sign reversal in the radial direction near the bar corotation should occur near the Lagrange points L1 and L2 along the bar's major axis. Since, inside the corotation radius, gas flows are directed towards the center of the galaxy associated with x1 orbits \citep[i.e. orbits along the bar's major axis, see Section 3.3 in][]{Binney_2008}. However, our approach traces radial velocities only near the minor axis of the disc, so by construction, the method is expected to work only when the bar major axis is aligned with the disc minor axis (PA close to 90°). Indeed, two out of three galaxies (id500577 and id506720) show agreement between the histogram peak and the true CR position at PA = 90°. For the third galaxy (id467415), the discrepancy at this orientation may be caused by stronger spiral arms (larger azimuthal velocities) and a tidal interaction that also enhances the vertical velocity component. More generally, we can conclude that the method suffers from strong orientation dependence and sensitivity to azimuthal and vertical velocities, which limit its applicability. Projection effects, the non-uniform azimuthal distribution of sign-change locations, and contamination from non-radial motions can all cause true peaks to vanish or false ones to appear. Thus, the method can yield CR estimates for specific orientations (PA close to 90°), but its strong orientation dependence and sensitivity to the contributions of azimuthal and vertical velocities limit its general applicability.
\par
We remind that this method has already been successfully tested on a bar model in~\citet{Font_2014a}. However, those tests considered only orientations where the bar was nearly aligned with the disc minor axis. Since theory predicts a sign change of the velocity at corotation near the bar major axis, it is not surprising that the pattern was clearly traceable near the disc minor axis in those cases. Moreover, the model used in those tests had spirals with weak amplitude.
\par
Although the results of applying this method to bar resonances are inconclusive, this does not exclude its applicability to estimating the spiral resonances. 
Besides, the negative outcomes may also be related to some issues with bar in TNG50. In this simulation bars are shorter than observed at the same angular velocity \citep{Roshan_2021, Frankel_2022}. Probably this discrepancy is related to the low mass of the stellar disc with the same mass of the dark matter halo, which in turn may be related to the subgrid physics of AGN feedback \citep{Fragkoudi_2021}. Finally, we cannot exclude the possibility that non-linear coupling between the bar and spiral arms near corotation could affect the gaseous kinematics in ways that complicate the interpretation.

\subsection{Kinematic methods: spiral nature}
\label{sec: results_2}
Here we test both methods described in Sections~\ref{sec: Canzian method} and~\ref{sec: FB method} to understand whether their application allows us to both identify the corotation position and determine the nature of the spiral arms. This analysis becomes more complicated because we are unable to reliably determine the corotation position of the spiral structure in the same way as was done for the bar. Moreover, we cannot clarify precisely which scenario of spiral formation actually takes place, even for simulated galaxies. 
\par
\subsubsection{Toy spiral galaxies}
\label{sec:toy models}
\begin{figure*}
\includegraphics[scale=0.1]{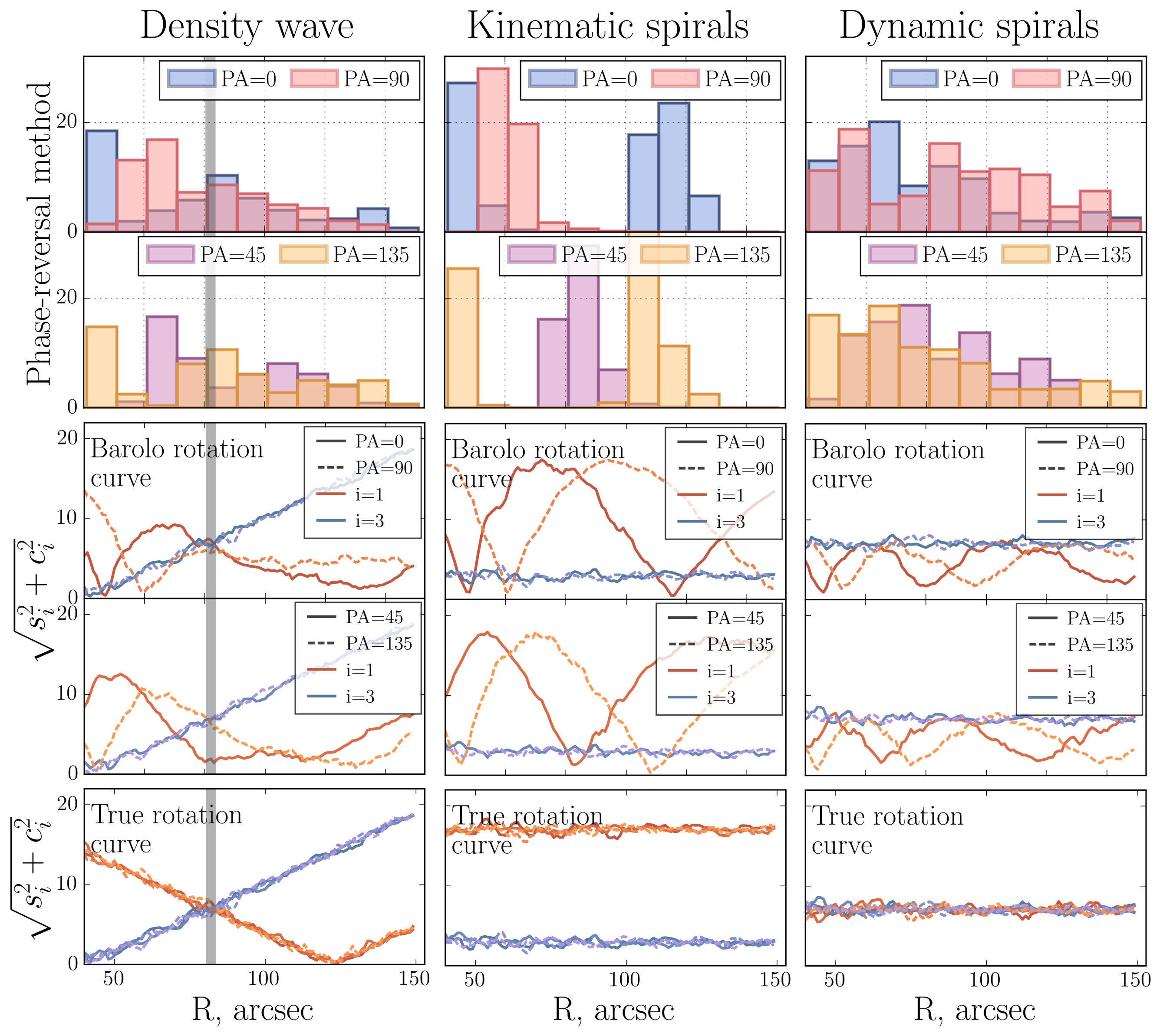}
\caption{Each column demonstrates the outcomes of the kinematic methods for toy models exhibiting density wave (left), kinematic (centre), and dynamic (right) spirals. The top two rows present histograms of ``phase reversal'' locations for different major axis orientations (see legend) at an inclination of $i = 30^\circ$. The third and fourth rows show the results of the other method applied with \textsc{Barolo} for different orientations (two per panel). The bottom row contains essentially the same for four orientations, but using the true rotation curves. The lines coloured with red correspond to the first-order velocity harmonic amplitude for different orientations, while the blue lines correspond to the third-order amplitude. The grey vertical line marks the CR position for the density wave case. Solid and dashed lines in the third and fourth rows refer to different PA orientations (see the legend on the corresponding panels).}
 \label{fig4}
\end{figure*}
We therefore begin with the simple toy galactic models with a flat rotation curve ($V_c=220$ km/s) before turning to the more realistic simulations. We created three toy models exhibiting two symmetrical arms corresponding to different scenarios: density waves $\Omega_\mathrm{p}=const$, kinematic spirals $\Omega_\mathrm{p}=\Omega - \kappa/2$, and dynamic spirals $\Omega_\mathrm{p}=\Omega$. To obtain at least a rough estimate of the reliability of these methods, we apply them to a mock velocity map. The map is generated under the assumption of linear theory directly using Equations~\ref{eq: c1s1c3s3}. We also add a small Gaussian noise in the same manner as described in Sec.~\ref{sec: data simulation}.
\par
Figure~\ref{fig4} demonstrates the results of both methods: the top rows correspond to the phase reversal method, while the two bottom rows belong to velocity harmonic dominance approach. Each column is associated with a distinct spiral nature: density wave (left), kinematic (middle), and dynamic (right) spirals. We first refer to the histograms of ``zero velocity'' positions presented in the top rows. As in the previous section, we considered several orientations of the position angle at a fixed inclination $i=30^\circ$, marked by different colors (see legend on each panel). Comparing the three cases related to the various toy models leads to some quite curious notes and conclusions. On the one hand, according to the left panel we do see a clear consistency between the position of the CR (grey vertical line) and the peaks of the histograms (though not for all orientations!) for the density wave model. On the other hand, much more noticeable peaks arise for the kinematic spirals case, for which by definition there is no corotation resonance. As for the dynamic spirals model, the histograms have no clear peaks and resemble a fence-like pattern.
\par
\begin{figure}
\centering\includegraphics[scale=0.5]{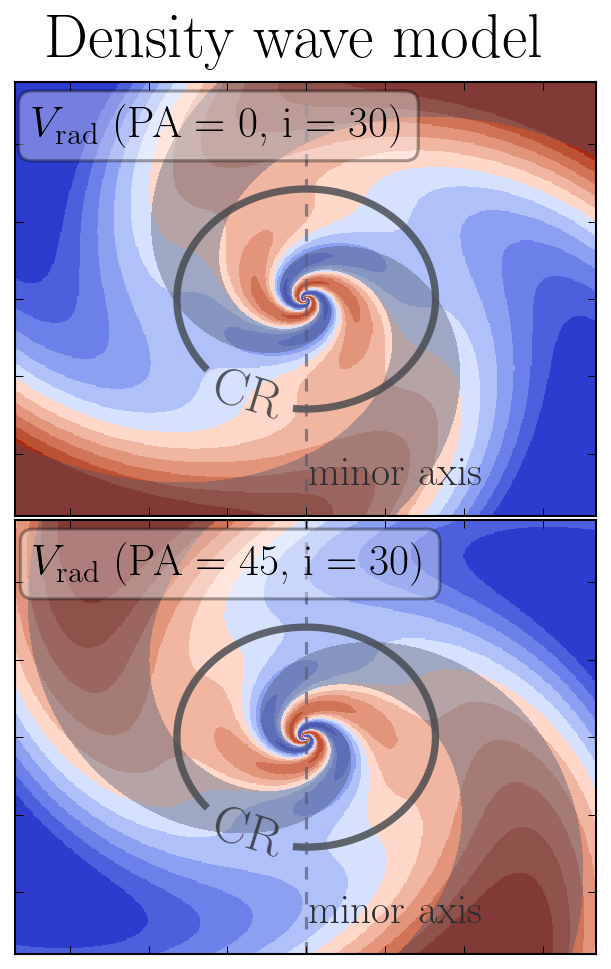}
\caption{Radial velocity distribution for the density wave toy model for two major axis orientations in the disc plane: PA = 0$^\circ$ (top) and PA = 45$^\circ$ (bottom). The grey ellipse marks the corotation radius. In the colour map, red corresponds to positive velocities and blue to negative velocities. The grey shaded areas indicate the spiral arms. }
 \label{fig5}
\end{figure}

It is worth returning to the density wave model to explain why for some orientations the peaks in the histogram do not coincide with the true CR location. We refer to Figure~\ref{fig5}, where radial velocity maps for orientations PA = $0^\circ$ and $45^\circ$ are shown in colour, with negative values in blue and positive ones in red. According to the top left panel of Figure~\ref{fig4}, for PA = $0^\circ$ (blue histogram) the phase reversal method recovers the precise corotation position. The top map of Figure~\ref{fig5} indeed shows a sign change of the radial velocity component in the radial direction near the minor axis in the vicinity of the CR (black circle). In contrast, the bottom distribution does not demonstrate a similar pattern near corotation and the minor axis, which apparently causes the discrepancy between the peak for the PA = $45^\circ$ orientation and the true resonance location (see purple histogram in Figure~\ref{fig4}). Paying attention to the grey shaded areas marking the spiral arm positions, we can conclude that the method works correctly only if the orientation of the disk is such that the spiral pattern occurs near the minor axis at the corotation circle. This imposes a strong restriction, since we do not know the true corotation position in advance. This constraint should be kept in mind during the analysis. Nethertheless, this approach can serve as the independend tool for estimating the CR position and for diagnosing the spiral origin.

In addition, we note that the histograms of the density wave and dynamical spirals have similar column heights. Considering more complex and realistic data will apparently add some additional noise to both distributions, and there is a possibility that the histograms for these cases may become difficult to distinguish. Thus, to avoid misinterpretation, we propose to use an additional independent approach, described in Section~\ref{sec: Canzian method} (see bottom panels of Figure~\ref{fig4}). The main idea of this method is based on the decomposition of the observed velocity field into a series (see eq.~\ref{eq: v_los}). For two-armed spirals, one should consider the first and the third terms for sine $s_{1,3}$ and cosine $c_{1,3}$ (see Section~\ref{sec: Canzian method}). As a reminder, according to linear theory for two symmetrical arms rotating as a rigid body, a change in the order of the amplitudes from the first to the third is expected at the corotation radius. This implies that the function $\sqrt{s_1^2(R)+c_1^2(R)}$ dominates over $\sqrt{s_3^2(R)+c_3^2(R)}$ inside the CR, and vice versa outside it.
\par
Since obtaining $c_1$ depends on the rotation curve, we examine two cases: one with the true $V_\mathrm{rot}(R)$ (bottom row) and one with the same profile extracted by \textsc{Barolo} (middle row). This allows us to estimate whether uncertainties in rotation curve can change the results and complicate the interpretation. In addition, we considered four different PA orientations, marked by different shades of red for the first terms and shades of blue for the third ones.
\par
The method applied to the density wave case demonstrates a pattern where the third harmonic amplitude rises with distance and becomes dominant over the first order at some radius. Note that such behaviour is observed for both the \textsc{Barolo} and true rotation curves, although in the former case, for some PA orientations, the phase change implying the CR is shifted toward the central regions. In contrast, for the dynamic spirals, the first and third harmonics have almost the same amplitude, which remains constant with radius. Thus, although this approach estimates the CR position inaccurately for some orientations, it allows us to definitively distinguish between dynamic and density wave spirals, unlike the other method. Moreover, according to the middle panel referring to the kinematic spirals, the behaviour of the harmonic amplitudes differs from that of the model with other spiral natures. The dominance of the first-order amplitude over the third-order one across the entire disk is typical for this model. This also holds for the case with the \textsc{Barolo} rotation curve, but with typical amplitude jumps in the first harmonic.
\par
Thus, we have demonstrated the main features that can help clarify the spiral nature of the galaxy under consideration and estimate the CR location in the case of a density wave presence. However, since these models are still too far from real galaxies, we now aim to examine the applicability of both methods using spiral galaxies in the TNG50 simulations. This analysis may reveal other limitations and caveats that we could have missed while investigating the simple models.

\subsubsection{Spiral galaxies from TNG50 simulation}
\begin{figure*}
\includegraphics[scale=0.12]{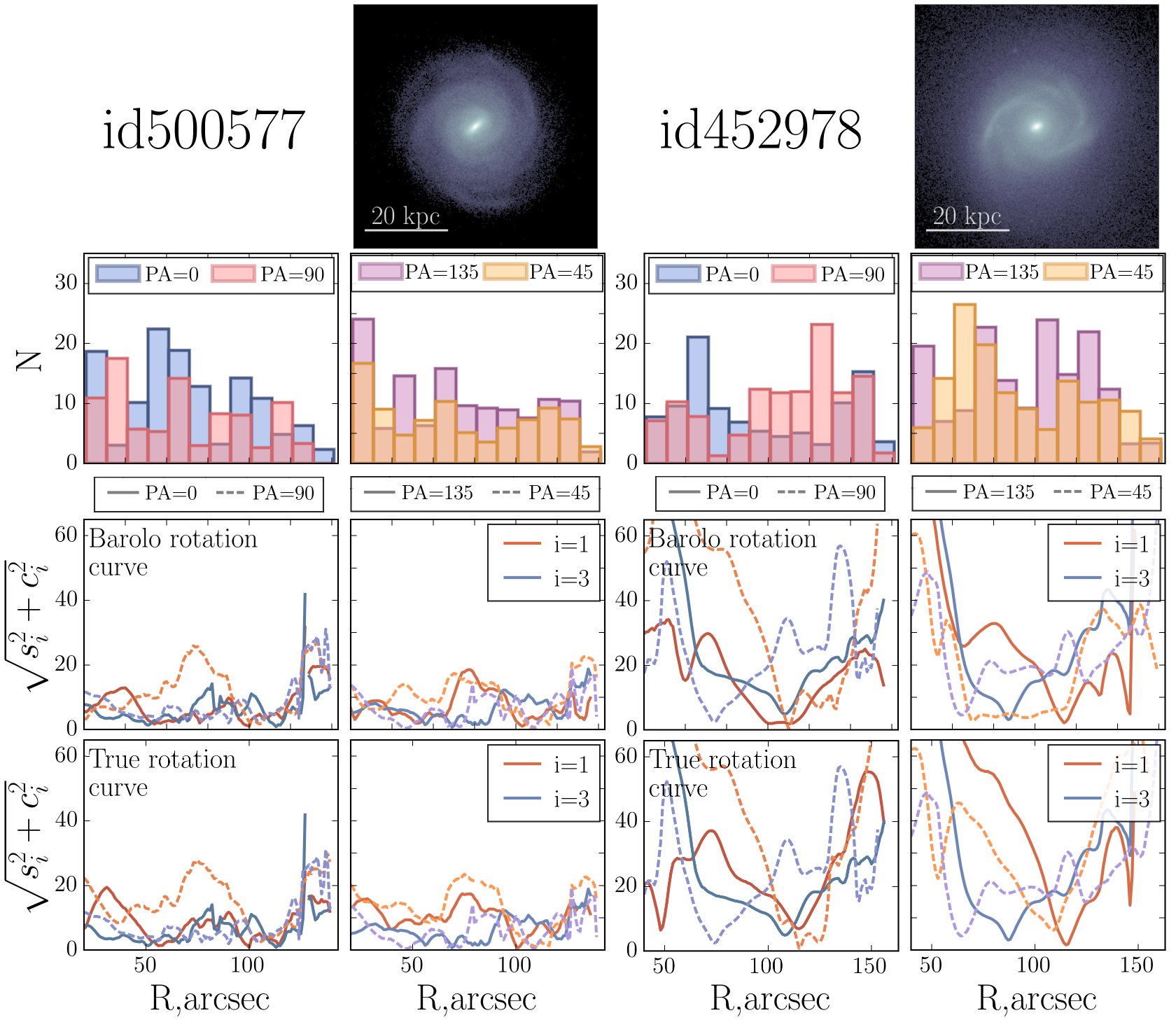}
\caption{Results of the kinematic methods as in Figure~\ref{fig4}, but with two columns corresponding to one model galaxy (the two left columns refer to id500577, the two right to id452978). Solid and dashed lines on the middle and bottom rows correspond to different PA orientations (see legend at the top of the central row). The stellar distribution map of the corresponding galaxy is shown at the top of each panel.}
 \label{fig6}
\end{figure*}

\begin{figure}
\includegraphics[scale=0.12]{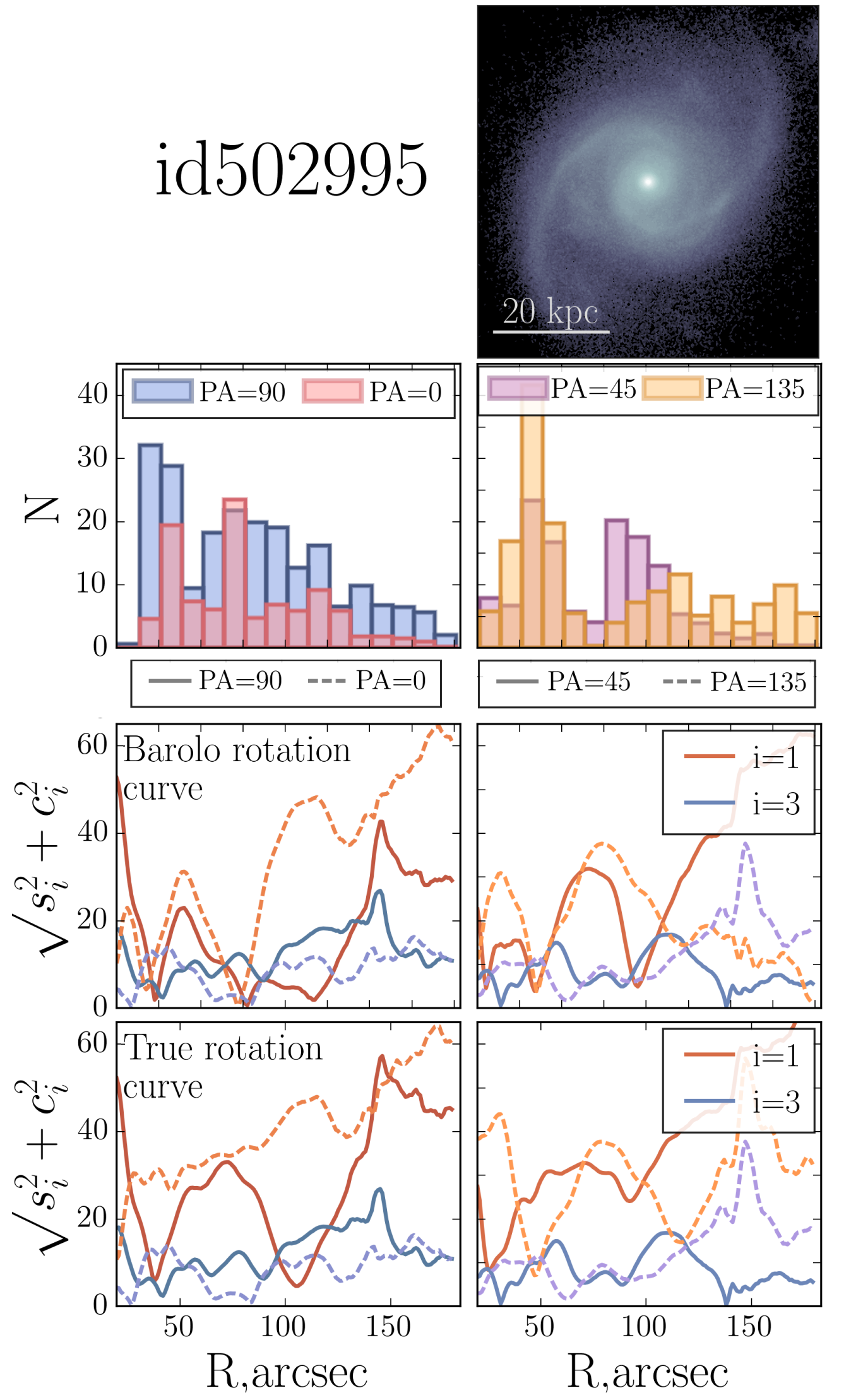}
\caption{The same as in Figure~\ref{fig6}, but for galaxy id502995.}
 \label{fig7}
\end{figure}

Figures~\ref{fig6} and~\ref{fig7} demonstrate the results of applying the kinematic methods to spiral galaxies from TNG50 (two columns for each object) considered at different positional angles and fixed inclination $i=45^\circ$. As examples, we chose three models with two-armed spirals, though they differ in several other aspects (Fig.~\ref{fig6} shows id500577 and id452978; Fig.~\ref{fig7} shows id502995). Model id500577 (considered earlier in Section~\ref{sec: results_1}) has a weak bar in the central region, and its spirals are less prominent than those in the other models. The strengths of the spirals in id452978 and id502995 are similar (see images of galaxies at the top of Fig.~\ref{fig6} and~\ref{fig7}); however, in the former case they are more tightly wound. The layout of both figures is similar to that of the simple toy models (see Figure~\ref{fig4}), except that each column contains information for two disc orientations (see legend in the top row) instead of four.
\par
As for the phase reversal method, the distributions of ``zero velocity'' distances (top row) change notably with orientation, especially for models with prominent spiral structure (id452978 and id502995). This approach does not allow us to make a statement about the spiral nature, since according to the results from the simple models (Section~\ref{sec:toy models}), even for objects with a clear CR position, this method is too sensitive to orientation variations. 
The method nevertheless remains useful for locating the corotation position, though within its known constraints. For a robust determination of the spiral nature, it should be used in tandem with the second kinematic method. As we showed for the toy models, the velocity harmonic dominance method complements the phase reversal results: it is independent, and it can unambiguously distinguish some spiral scenarios that the phase-reversal method cannot.
\par
Considering velocity harmonic dominance method applied to more realistic models, we first note that the radial profiles of the first- and third-order harmonic amplitudes generally behave similarly for the \textsc{Barolo} (middle row) and true (bottom row) rotation curves. However, at some PA for id452978 and id502995 we observe differences in the amplitude of the first harmonic, which arise at the same distances where strong deviations of the ``observed'' rotation curve from the true one occur (see Fig.~\ref{apfig3}). Note that these exact variations of the first-order harmonic can lead to a shift in the harmonic crossing. For example, noticeable differences between the mutual distribution of the first- and third-order harmonic amplitudes are observed in the central region of id452978 at PA=135, where, according to Fig.~\ref{apfig3} (left), the \textsc{Barolo} rotation curve is strongly underestimated due to the almost empty gas distribution along the major axis. Nevertheless, we conclude that using an inaccurate $V_\mathrm{rot}$ radial profile does not strongly misrepresent the results.
\par
Secondly, it is easy to notice a much more significant dependence on orientation that was not typical for the toy models. Since this holds for the case with the true rotation curve, such variations are not caused by uncertainties in the $V_\mathrm{rot}(R)$. One of the reasons for such discrepancies may be related to a physical change in the ionised gas content along the line of sight with orientation. Note that the deviations that are significant for the interpretation correspond to models with strong spiral structure, especially id452978. Additional analysis reveals that id452978 has a relatively large magnitude of the vertical velocity component $V_\mathrm{z}$, which also causes significant changes in the residual velocity distribution and, apparently, in the amplitude of the velocity harmonics. Thus, a small contribution of $V_\mathrm{z}$ to the observed velocity is one of the limitations of the considered approach. One more feature we observe is a relatively small amplitude of the velocity harmonics for model id500577. This is explained by the lower magnitudes of the radial and azimuthal non-circular velocities (directly related to $c_1$ and $s_1$), which is typical for galaxies with less pronounced spiral structure. Taken together, these findings indicate that the discrepancies between both the harmonic crossings and the phase-reversal peaks at different PA cannot be attributed to rotation curve uncertainties alone: a physical dependence of the line-of-sight gas kinematics on the disc orientation, as well as the contribution of the non-circular azimuthal and vertical velocity component, likely play a role as well.
\par
As for the identification of the spiral nature, the analysis of the velocity harmonic profiles does not reveal any hint of a density wave for any of the simulated galaxies. It is very probable that model id502995 has kinematic spirals (see Figure~\ref{fig7}). First, the histograms in the top row for some orientations have very clear peaks, which was typical for the toy model of the same nature. Second, the amplitude of the first term (red lines) is generally higher than that of the third term (blue lines). Moreover, noticeable jumps in the amplitude of the first harmonic are present, as also noted previously. An examination of the simulation itself shows that there was an interaction with a companion, which apparently induced the formation of the two-armed structure in this galaxy. This confirms that, at least for kinematic spirals, the method could be robust.
\par
Considering the results for id500577 (Figure~\ref{fig5}, left), both the ``phase reversal'' distributions and the velocity harmonic amplitudes are similar to those presented for the dynamic spiral toy model (Figure~\ref{fig6}, right). Specifically, we see fence-like histograms (i.e., many peaks of comparable height), with the exception of a few peaks likely related to the bar resonances (see middle row of Fig.~\ref{fig3}), and similar amplitudes of the first and third terms, which are almost constant with radius. In contrast, the id452978 model (Figure~\ref{fig6}, left) shows results that do not resemble any of those obtained for the toy models. During the toy model analysis, we were unable to consider one additional important scenario: the existence of several modes simultaneously. It is possible that the id452978 model corresponds to exactly this case. In addition, \citet{Schoenmakers_1997} suggested that strong variations in the $s_3$ and $c_3$ terms may indicate several co-existing spiral modes.
Unfortunately, the methods we have cannot independently clarify the nature of the spirals in the simulations (except for the tidally induced ones, id502995). Therefore, the suggestions presented above for id500577 and id452978 remain hypotheses.
\par
Thus, before moving on to applying the method to real data, we will summarize all limitations we revealed during the analysis. Both methods share similar physical and geometric constraints: the gas content along the line-of-sight, strong non-circular azimuthal and vertical velocity contributions, and the accuracy of the input rotation curve can all shift peaks or crossings from the true ones. The rotation curve uncertainty is the most tractable of these, since we can identify the regions where the extracted curve is least reliable, such as areas with sparse gas or a strong spiral contribution along the major axis.

\subsection{Application to the observed spiral galaxy}
\label{sec: application}
\begin{figure*}
\includegraphics[scale=0.55]{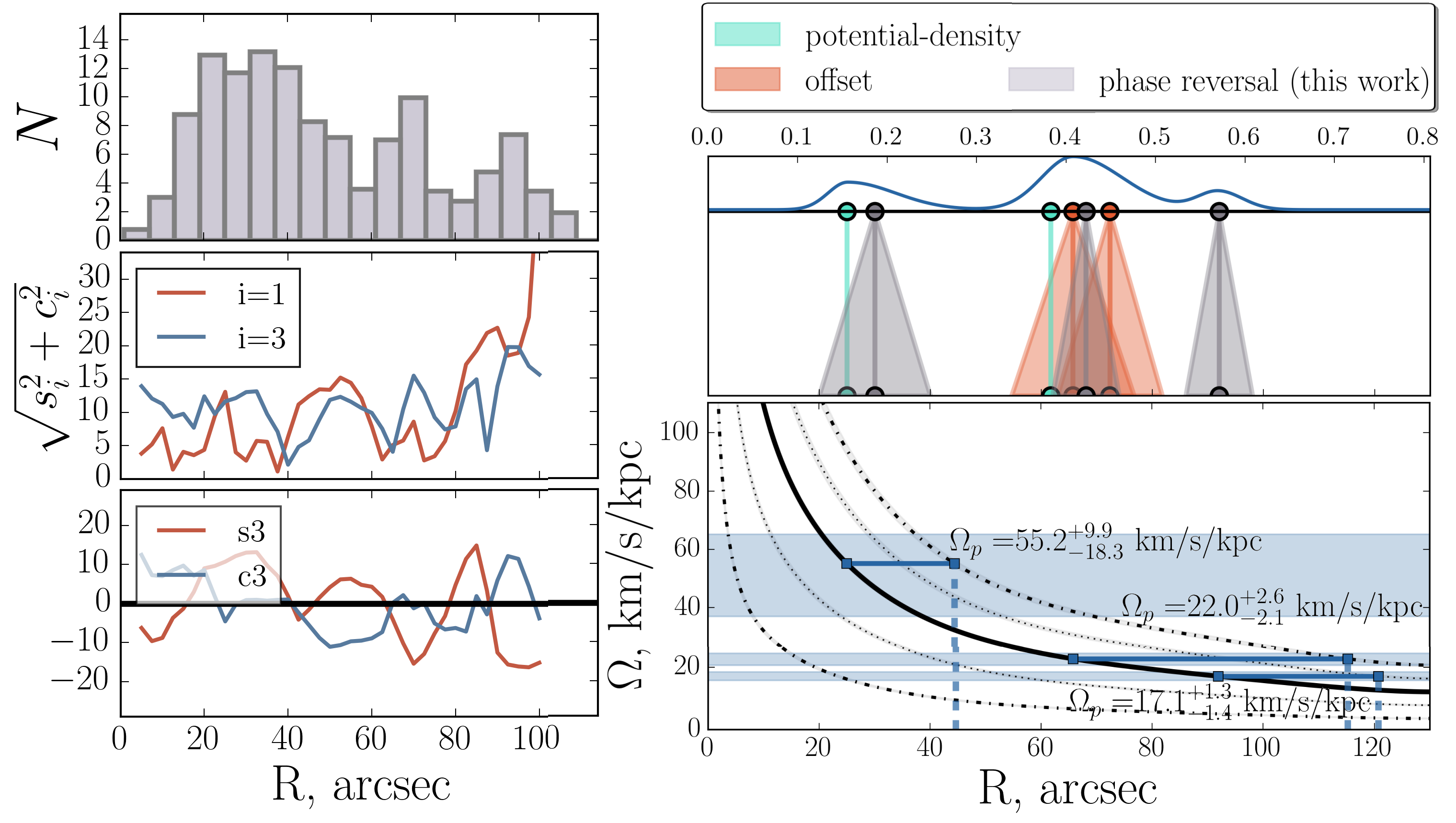}
\caption{Left panel: results of kinematic methods applied to H$\alpha$ data of NGC~3938  — distributions of ``zero velocity'' distances (top), radial profiles of the first- and third-order harmonic amplitudes (middle), and the 
$s_3$ and $c_3$ terms (bottom). Right panel: distribution of corotation radius measurements taken from the literature (red and cyan) and estimated in this paper (grey); the shaded areas indicate the error of each measurement (the base of the cone), and their colour indicates the method used. The blue line at the top shows the fitted asymmetric Gaussian function. The black solid line in the bottom panel shows the angular velocity profile $\Omega(R)$ calculated from the rotation curve extracted using \textsc{Barolo}; dashed lines to the left and right mark the locations of the inner and outer Lindblad resonances, and ultraharmonic resonances are plotted with dotted lines. The blue horizontal lines connect the positions of the corotation radius and the OLR or oUHR for the corresponding pattern speeds.}
 \label{fig8}
\end{figure*}

Figure~\ref{fig8} (left panel) presents the results of the velocity dominance and phase reversal methods obtained using the H$\alpha$ residual velocity map of the observed spiral galaxy NGC~3938 (see Figure~\ref{fig2}). Although the distribution of ``zero velocity'' distances (top left) appears rather noisy, it contains three notable peaks that could indicate the corotation radius: a broad one at 20–40 arcsec and narrower ones at about 70 and 90 arcsec. In previous sections we found that the phase reversal method strongly depends on the galactic disc orientation, and it is impossible to determine the correct orientation in advance. Moreover, the amplitudes of the azimuthal and vertical non-circular velocities can also affect the resonance position measurements. The low inclination of NGC~3938 ($\sim 10^\circ$) is intrinsically unfavourable for line-of-sight kinematic methods such as those considered here. At small inclinations, the projection of the vertical velocity component onto the line of sight is not suppressed, so vertical motions can contribute more strongly than in more inclined discs. However, the vertical velocities in this galaxy are likely small: there are no clear signs of interaction, the velocity field and the residual velocity map appear fairly regular. We can therefore assume that the contribution of $V_\mathrm{z}$ does not strongly affect the results.
\par 
According to dataset of corotation radius measurements~\footnote{\url{https://github.com/ValerieKostiuk/CRs\_dataset}} gathered for about 500 galaxies~\citep{Kostiuk_2024}, nine CR positions for this galaxy have been reported in the literature using four different approaches. We excluded five measurements from consideration: three were marked by the authors as unreliable~\citep{MartinezGarcia_2009, Meidt_2009}, and two (from~\citealp{Elmegreen_1992, Elmegreen_1995}) were obtained using morphological features and can only be interpreted as additional evidence. The remaining four measurements consist of two values from~\citet{Abdeen_2020} (shown as red cones on the top right panel of Fig.~\ref{fig8}) and two from~\citet{Buta_2009} (cyan vertical lines). 
\par
On the top right panel of Fig.~\ref{fig8} we plot the three CR estimates obtained from the phase-reversal method as grey cones. Comparison with other CR positions obtained by independent methods \citep{Buta_2009, Abdeen_2020} shows agreement with our two inner measurements, strengthening our confidence in both the method and the results. We estimated the most likely CR position by fitting the distribution of measurements with an asymmetric Gaussian function (shown by the blue line at the top). The uncertainties were calculated as the standard deviation of the fitted distribution divided by the square root of the number of measurements. Thus, we identified three CR positions at 
 $25_{-3}^{+8}$, $64_{-4}^{+6}$ and $91_{-5}^{+6}$ arcsec, suggesting the possible existence of at least three spiral modes rotating at different pattern speeds (marked at the bottom-right plot). 
\par
The results of the velocity dominance method are shown in the middle row of Fig.~\ref{fig8}. From the mutual behaviour of the first- (red) and third-order (blue) harmonic amplitude profiles, we conclude that neither a single density wave nor a kinematic spiral scenario is relevant for this galaxy. This follows from the fact that we do not observe a clear change of prevalence from the first to the third harmonic, or the dominance of the first harmonic alone across the entire disc, as found in Section~\ref{sec: results_2} (see Fig.~\ref{fig4}). Notable variations in the third-order amplitude, which are not typical for the dynamic spiral scenario, also exclude that mechanism from consideration. Thus, although this method does not point to a specific scenario, it allows us to rule out those several possibilities. The three harmonic crossings that approximately coincide with the peaks of the phase-reversal histogram may indicate the presence of three spiral modes, but this interpretation requires further verification. Furthermore, pronounced changes of $s_3$ and $c_3$ terms (bottom left panel) may also point to a recurrent mode spiral nature.
\par
The pattern speeds of the modes related to the most likely CR position are marked on the bottom right panel by blue horizontal lines and shaded areas around them. The minimum and maximum angular velocity values of each pattern were set equal to the angular velocity of the disc (calculated from the rotation curve obtained with \textsc{Barolo}) at the outer and inner estimates of the corotation radius, respectively. It is worth noting the alignment of the OLR position associated with the first mode and the inner ultraharmonic resonance of the second mode. Furthermore, we observe the agreement of the outer ultraharmonic resonance of the second mode with the location of the corotation radius of the third mode. As known from theoretical studies \citep{Masset_1997} and numerical simulations \citep{Rautiainen_1999}, such an overlap provides an effective transfer of energy and angular momentum, which is necessary for spirals to survive long. Note also that for NGC~3938 we observe exactly the situation described in \citet{Marchuk_2024b}: resonance coupling occurs simultaneously for two pairs of resonances, specifically the OLR of the inner pattern coincides with the IUHR of the next, and the IUHR of the inner pattern broadly coincides with the ILR of the second. The ratio of the corotation radii in Figure~\ref{fig8} is about 2.5, as predicted for this case (see Table~1 in aforementioned work). This consistency of resonance positions also indirectly supports the recurrent spiral modes scenario.

\section{Discussion and Conclusion}
\label{sec:dicussion}

Kinematic data have been used repeatedly to clarify the nature of spiral arms in observed galaxies. The most popular approach is the Tremaine–Weinberg technique, which allows one to estimate the spiral pattern speed under the assumption that it is constant with radius~\citep{Tremaine_1984}. Although this method is based on rough initial assumptions and has limitations \citep{Borodina_2023}, it has been widely applied to both nearby galaxies \citep{Fathi_2009,Cuomo_2020,Williams_2021} and those at high redshift \citep{Roshan_2025}.
\par
The other methods we are mostly interested in focus on kinematic features that are expected in density wave spirals. \citet{Canzian_Allen_1997} were able to detect the radial change in velocity harmonic dominance for NGC~4321 that implies spiral density wave. Based on the same theoretical basis, \citet{Font_2011} approached this problem by searching for kinematical features occurring near resonances. This strategy enabled the examination of a large number of objects and demonstrated that the co‑existence of several spiral modes may be the dominant mechanism of spiral arm formation \citep{Font_2014a}. In addition, the results of this method were shown to be consistent with other corotation measurements and to be independent of which gas tracer was used \citep{Beckman_2018}.
\par
Our own strategy relies not only on examining kinematic features but also on combining as many independent observational tests as possible to confirm the spiral formation scenario for a given galaxy. First, we base our analysis on measurements of corotation radii determined by various methods reported in the literature \citep{Kostiuk_2024}. Second, we use corotation estimates obtained from other independent techniques. Furthermore, we examine specific kinematic and morphological features expected in different scenarios. Taken together, all these pieces of evidence allow us to robustly determine the spiral nature for each object. We conducted a similar examination for three morphologically distinct galaxies in \citet{Kostiuk_2025}; however, before drawing firm conclusions about the spiral nature for a larger sample, we need to verify the reliability of the adopted methods. Therefore, in this paper we focused on testing two kinematic methods using toy models and TNG50 simulations. The use of controlled simulations enables us to check both the technical implementation of the methods and the basic assumptions implied by theory.
\par
Before investigating spiral arms, we tested the reliability of the phase reversal method for extracting bar resonance locations using three models from the TNG50 simulations. First, we found that the radial velocity component traces the CR position better than the purely non-circular velocity map (see Appendix~\ref{sec: initial assumption}). Note that we did not find any signs of the expected features near the other resonances (i.e., OLR, oUHR). Second, we revealed that the accuracy of the method strongly depends on the contribution of vertical and azimuthal velocity components. To minimise the vertical component influence, one can consider highly inclined disc galaxies or those with different inclinations that possess a dynamically cold disc; to reduce the impact of non-circular azimuthal motions, excluding galaxies with strong spiral arms may help. Moreover, since radial velocities can be accurately estimated only near the minor axis of the disc, this approach is restricted by the bar orientation in the galactic plane. This limits the sample to those galaxies for which the major axis of the bar coincides with the minor axis of the disc.  Although this method has been applied to measure bar resonances and pattern speeds in a large number of galaxies~\citep{Font_2014b, Font_2019} and was shown to be robust~\citep{Font_2014a}, our analysis has revealed important constraints that were apparently missed previously, likely due to a limited range of orientation parameters. These findings do not necessarily invalidate earlier results but suggest that they may be reliable only for galaxies with favourable orientations.
\par
We also explored an alternative approach based on a sign change of the $s_1$ harmonic, which has been used in some studies to locate the bar corotation \citep[e.g.,][]{Salak_2019}. For some of our barred galaxy models, the sign reversal of $s_1$ did coincide with the true CR position, but for others no clear sign change was observed at any orientation. Because of this ambiguity, we chose not to include these results in the main body of the paper.
\par
Returning to the question of spiral nature, in Section~\ref{sec: results_2} we examined the results of both kinematic methods on a sample of model spiral galaxies. In previous investigations, these approaches could only provide signatures of a structure rotating at constant angular velocity $\Omega_\mathrm{p}$ since they focused on features predicted by density wave theory. However, recent cosmological simulations \citep{Quinn_2026, Grand_2026} have revealed that what we understand as spiral nature can change during a galaxy's lifetime — driven by tidal interactions, bar episodes, or simply evolution. This finding suggests that focusing on a single mechanism may be insufficient. For this reason, we instead aimed to obtain attributes typical not only for density wave, but for kinematic and dynamic spiral scenarios. Using simple toy models (see Fig.~\ref{fig4}), we found that the phase reversal method cannot unambiguously determine either the CR position or the spiral nature: the distribution of ``zero velocity'' distances shows clear peaks for the kinematic spiral scenario, while for the density wave case the peaks are less clear and could be misinterpreted as dynamic spirals. The analysis of velocity harmonic amplitudes resolved this discrepancy: dominance of the first-order amplitude over the third-order one is typical for kinematic spirals, equal amplitudes of these harmonics correspond to dynamic spirals, while in the density wave case there is a change in dominance at corotation. Thus, we suggest that both kinematic methods should be applied as complements to one another in order to obtain a reliable prediction of the spiral origin.
\par
The application of these methods to more realistic modelled spiral galaxies from the TNG50 simulations complicates the interpretation of the results (see Fig.~\ref{fig6} and Fig.~\ref{fig7}). These complexities likely arise from several factors, such as non-perfectly symmetric arms and not tightly wound spirals — features that cannot be avoided in reality. However, the most significant effect is caused by a high contribution of the vertical velocity component, which makes the methods strongly sensitive to orientation. An additional difficulty is our inability to clarify the exact spiral scenario using robust approaches — for example, the analysis of spectrograms of the dominant Fourier component, which essentially shows the radial distribution of the pattern, as done in the \textsc{Auriga Superstars} \citep{Grand_2026} and FIRE-2 \citep{Quinn_2026} simulations. The issue is that such spectrograms can only be extracted from models with high temporal resolution, which is lacking in TNG50.
\par
One more important finding of our analysis is that the results are not strongly dependent on uncertainties in the rotation curve. The extraction of a precise $V_\mathrm{rot}(R)$ from observations of disc galaxies can be crucial for some problems. Over the past three decades, numerous software packages have been developed; the most recent include \textsc{Gbkfit} \citep{Bekiaris_2016}, \textsc{2DBAT} \citep{Oh_2018}, \textsc{MCMC} \citep{Oh_2019}, and \textsc{VRMA} \citep{Labini_2023}. In this study, we chose \textsc{Barolo} (2D) because it is simple to use and sufficiently fast (compared to \textsc{Barolo} 3D). However, as demonstrated in several studies \citep[e.g.,][]{Teodoro_2015, Salibur_2026}, the rotation curves derived with this technique can differ significantly from the true one, especially for barred galaxies, even beyond the bar region. We have shown that using an inaccurate profile may in some cases alter the results, but it does not complicate the interpretation. The contribution of these uncertainties turns out to be less critical compared to the other, more important factors discussed above.
\par
Despite these limitations, the kinematic methods considered can still be useful for clarifying the spiral nature, since the outcomes of both methods show patterns similar to those obtained from the simple toy models. Moreover, this analysis has given us an idea of what the results would look like when applied to observed velocity fields. In Section~\ref{sec: application} we examined the question of spiral nature in NGC~3938 based on the results of both kinematic methods. We estimated the corotation radii using the phase-reversal approach and compared them with literature values. This analysis revealed several corotation resonance locations, indicating a recurrent spiral modes scenario. The consideration of the third-order velocity harmonics and the overlap of resonances further supports this interpretation.
\par
Also, kinematic methods are not the only ones which can be used to determine the nature of spiral structure \citep[see an overview of different methods in][]{Kostiuk_2024}. In the next works, we intend to extend our analysis to photometric methods in the same manner, applying existing techniques to the data from simulations (including the same sample of galaxies from TNG50) in order to validate them and identify their strong and weak sides.
\par
Thus, we have demonstrated a strategy that enables a valid prediction of the spiral structure formation mechanism for observed galaxies. Applying this analysis to a larger sample of galaxies and using wider range of methods will help to uncover the diversity of spiral formation scenarios and to assess the prevalence of each mechanism in the local Universe.

\section*{Acknowledgements}

The authors thank Alexei V. Moiseev (SAO RAS) for consultations on FPI BTA observations and data reduction, and Dmitry V. Oparin (SAO RAS) who performed the BTA observations. This study was supported by the Russian Science Foundation, grant number 25-22-00543. Observations with the SAO RAS telescopes are supported by the Ministry of Science and Higher Education of the Russian Federation. The renovation of telescope equipment is provided within the national project 'Science and Universities'. V.K. thanks the Fund for the Development of Theoretical Physics and Mathematics 'BAZIS' grant number 23-2-2-6-1.
\par
We also acknowledge the anonymous referee for his/her thorough review and highly appreciate the very useful comments and suggestions that contributed to improving the quality of the article.

\section*{Data Availability}

The data underlying this article will be shared on reasonable request to the corresponding author.



\bibliographystyle{mnras}
\bibliography{kostiukvs}




\appendix

\section{Test of phase reversal method initial assumptions}
\label{sec: initial assumption}

In this section, we aim to check the key idea of the method: corotation radius (or other resonances) can be detected by tracing the phase changing in non-circular velocity map. First, we need to rule out the possibility that the main assumption is incorrect, namely that zero non-circular velocity positions associated with sign changes of the velocity actually correspond to resonance locations. Using the procedure described in Section~\ref{sec: data simulation}, we obtain a map of non-circular velocities for the face-on orientation and apply the ``zero velocity'' detection algorithm to it. Note that this map contains the magnitude of the sum of the radial and azimuthal velocity vectors, excluding the vertical component. 

\begin{figure*}
\includegraphics[scale=0.12]{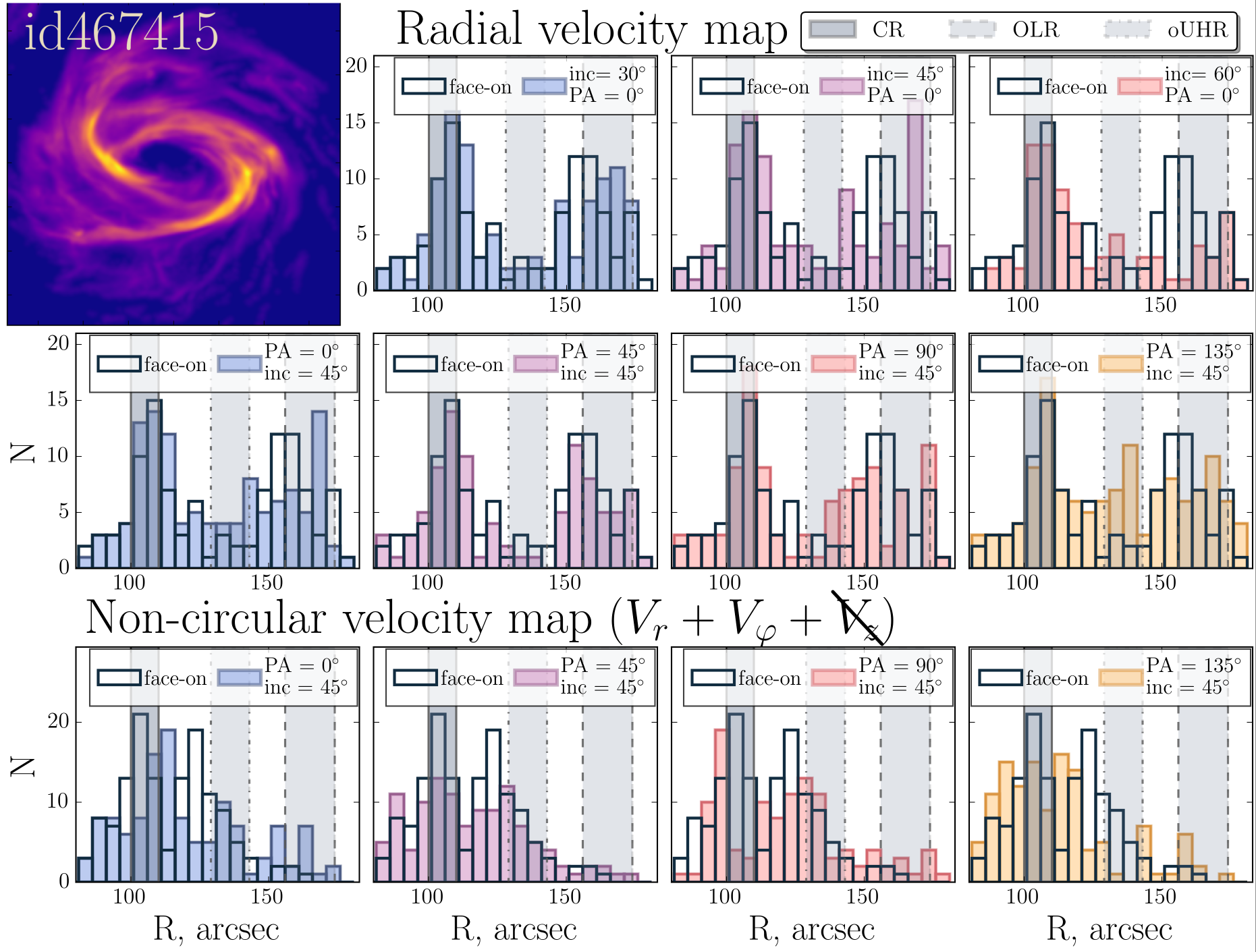}
\caption{Top and middle rows show the distributions of ``zero velocity'' distances obtained from the radial velocity map for inclination and PA variations, respectively. Coloured histograms refer to the inclined disc (see legend) and the black histogram to the face-on orientation. The bottom row shows the same but extracted from the non-circular velocity map (excluding the vertical component). Shaded vertical bands indicate the true positions of the main bar resonances (see legend at the top right). The top left panel shows the ionised gas distribution for the simulated galaxy considered.}
\label{apfig1}
\end{figure*}

The histogram outlined in black in the bottom panel of Figure~\ref{apfig1} shows the distribution of galactocentric distances at which the true radial velocity changes sign along the radial direction. It is easy to see that one peak in the distribution clearly coincides with the bar corotation radius, marked by the shaded area with solid vertical lines. As for the other resonances (OLR and oUHR), there is no explicit consistency between their positions and the peaks in the histogram. Although the vicinity of the outer ultraharmonic resonance (shaded area with dash-dotted vertical lines) definitely contains regions where changes in the radial velocity sign occur, the position of the corresponding peak is shifted from the true location. Therefore, we can conclude that sign changes of the velocity can indeed be traced near the resonances, but the locations of the main peaks in the histogram may misrepresent the true resonance positions.

We further investigate whether changes in the orientation of the disc relative to the observer can affect the results. We repeat the same procedure for non-circular velocity maps with varying position angles (PA = 0°, 45°, 90°, 135°) of the major axis at a fixed inclination of $i = 45^\circ$ (see coloured histograms in the bottom row). Note that PA = 0° and PA = 90° correspond to directions aligned with and perpendicular to the major axis of the bar and the disc, respectively. The non-circular velocity maps in this case represent the projected velocity values onto the line of sight.

Comparing the histograms for inclined discs with that obtained for the face-on orientation, we observe significant differences. For some cases (PA = 0° and PA = 90°) the true CR position can be estimated using these histograms, although with a shifted location. For other orientations, the level of the peaks is comparable to the noise level, thus complicating the identification of the corotation radius.

In contrast to the non-circular velocity maps, applying the same method to pure radial velocity distributions yielded more positive results (see the top and middle rows of Figure~\ref{apfig1}). The distribution of ``zero'' radial velocity for the face-on orientation, presented by the black-outlined histogram, shows consistency between the peaks and the positions of the corotation and OLR resonances. The coloured histograms in the top panel correspond to different inclinations at a fixed position angle PA = 0°, while the middle panel shows the ``phase reversal'' locations for different position angles at a fixed inclination of $i = 45^\circ$. According to these figures, all cases exhibit essentially the same peak at the bar corotation radius. As for the other main resonances, some peaks coincide with the true OLR and even oUHR positions; however, we cannot argue that this method can reliably predict their locations. First, for some disc orientations these peaks appear at different locations compared to those in the face-on case. Second, for certain orientations these features near resonances disappear entirely. Thus, although the CR position remains detectable across different orientations, projection effects can introduce false peaks in the ``zero velocity'' radial distribution, which could be misinterpreted as real resonances.

Therefore, based on this analysis, we conclude that applying the phase reversal method using the radial velocity distribution instead of the non-circular one yields more robust results. We have demonstrated that the use of the radial velocity map is less sensitive to orientation changes. Furthermore, using the radial component has a physical basis, since resonance positions are usually associated with radial gas flows, which are directly related to sign changes in the corresponding velocity component in radial direction.

\section{Comparison of morphology and velocity curves at different orientations}
\label{sec: different orientations}

\begin{figure*}
\includegraphics[scale=0.08]{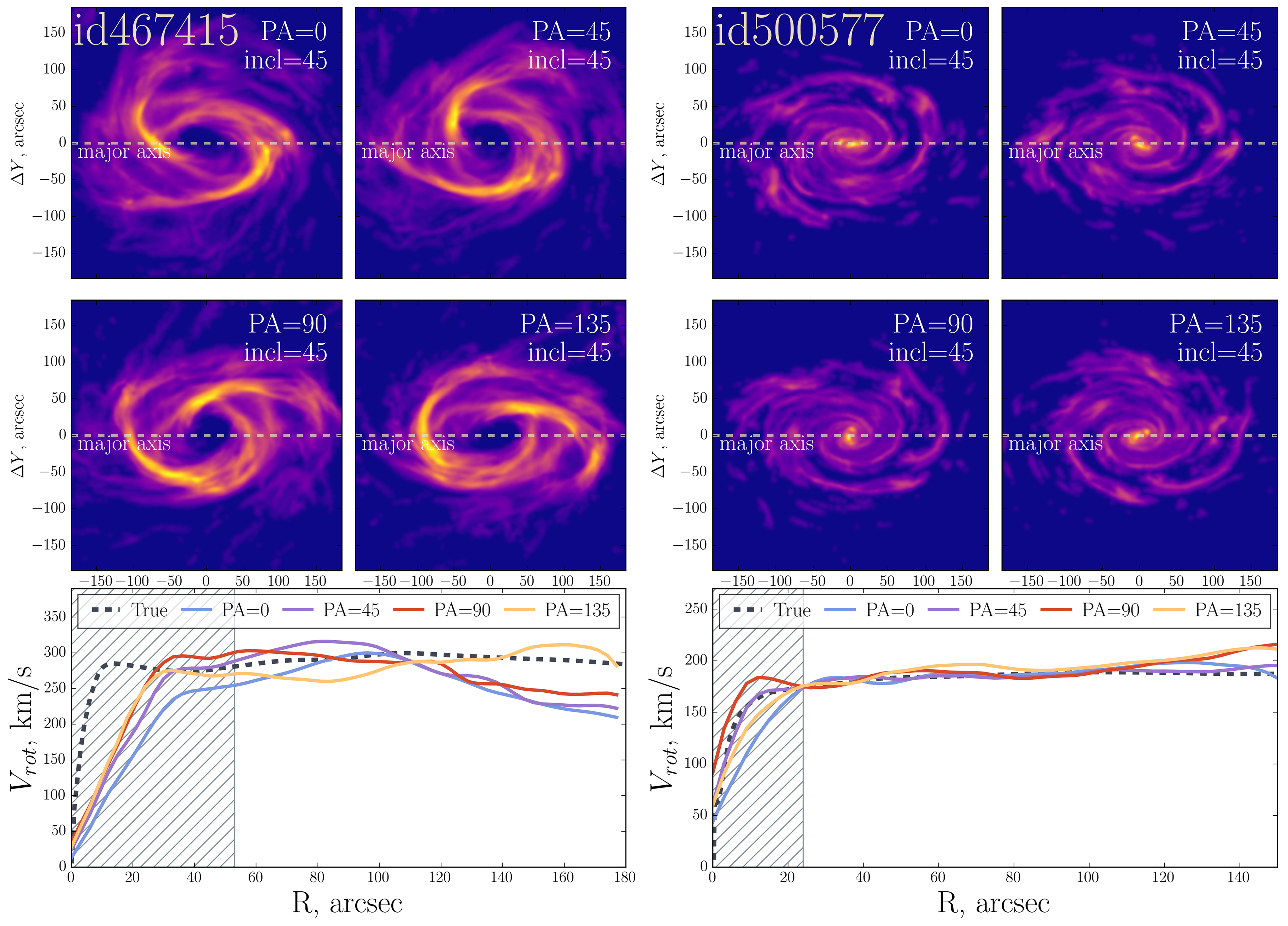}
\caption{The H$\alpha$ intensity maps for different orientations are shown in the top and middle rows. The bottom row shows the rotation curve extracted using \textsc{agama} (black dashed lines) and \textsc{Barolo} (coloured solid lines). The PA values are indicated in the legend at the top. The left panel corresponds to the simulated galaxy id467415, while the right panel corresponds to id500577. Hatched areas mark the bar region.}
\label{apfig2}
\end{figure*}

\begin{figure*}
\includegraphics[scale=0.08]{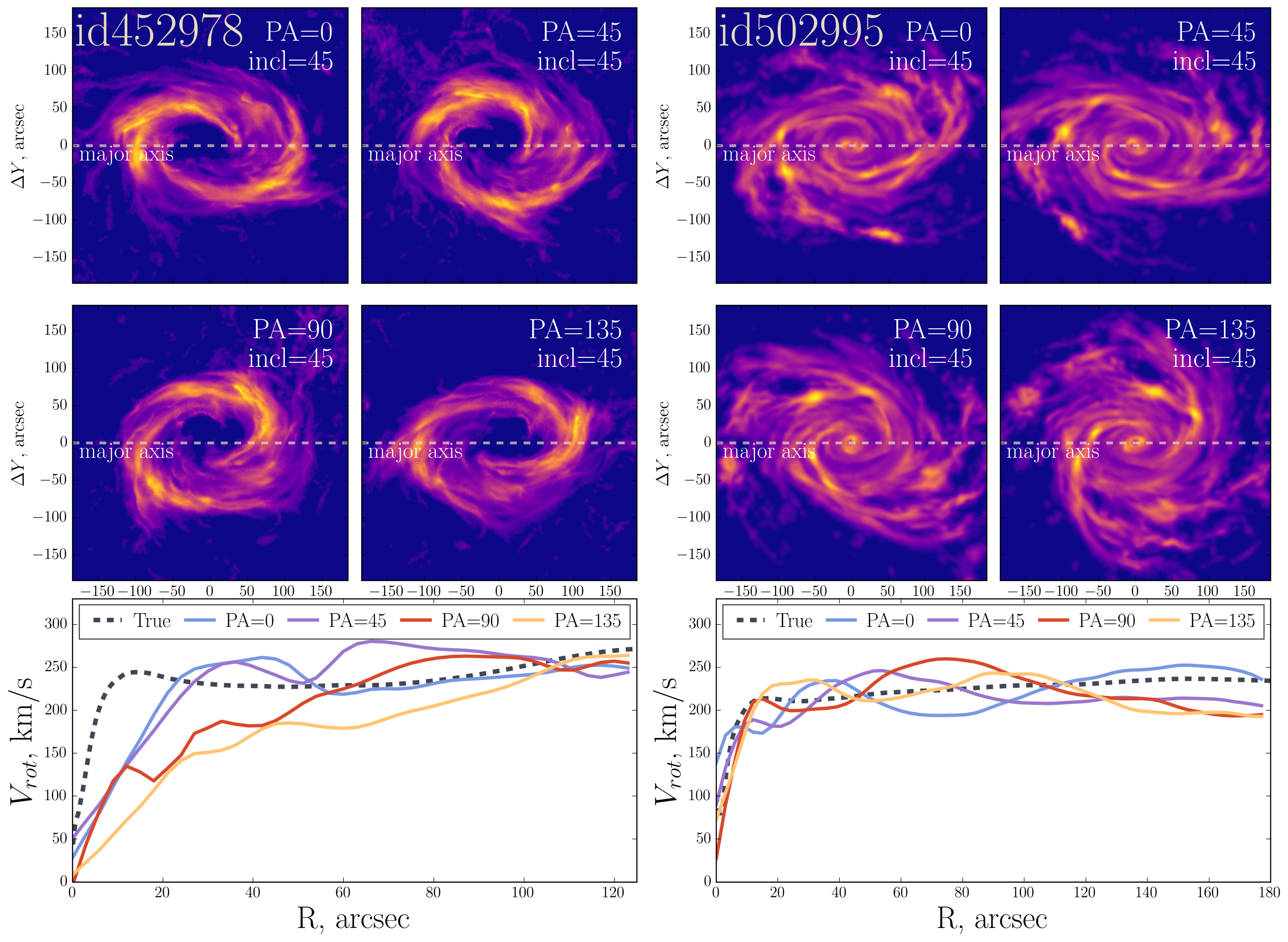}
\caption{The same as on Fig.~\ref{apfig3} but for unbarred galaxies id502995  and id452978.}
\label{apfig3}
\end{figure*}

\bsp	
\label{lastpage}
\end{document}